\documentclass[letterpaper,english,aps,prl,twocolumn,superscriptaddress,floatfix]{revtex4-2}
\usepackage[LGR,T1]{fontenc}
\usepackage{color}
\usepackage{babel}
\usepackage{mathrsfs}
\usepackage{amsmath}
\usepackage{amssymb}
\usepackage{graphicx}
\usepackage{verbatim}
\usepackage[pdfusetitle,
 bookmarks=true,bookmarksnumbered=true,bookmarksopen=true,bookmarksopenlevel=3,
 breaklinks=true,pdfborder={0 0 0},pdfborderstyle={},backref=false,colorlinks=true]
 {hyperref}
 \usepackage[normalem]{ulem}

\makeatletter

\pdfpageheight\paperheight
\pdfpagewidth\paperwidth

\usepackage{mathalpha}
\usepackage{xcolor}
\usepackage{simpler-wick}
\usepackage{footnote}
\usepackage{url}
\usepackage{mathtools}

\def\e{\textrm{e}}
\def\i{\textrm{i}}

\def\d{\textrm{d}}

\makeatother

\begin{document}
\title{Vacuum-Triggered Instability in Paired Superradiance}
\author{Zhan Bai}
\affiliation{State Key Laboratory of Ultraintense Laser Science and Technology, Shanghai Institute of Optics and Fine Mechanics, Chinese Academy of Sciences, China}

\author{Ningqiang Song}
\affiliation{Institute of Theoretical Physics, Chinese Academy of Sciences, China}

\author{Min Chen}
\affiliation{State Key Laboratory of Dark Matter Physics, Shanghai Jiao Tong University, China}

\author{Xiangyan An}
\affiliation{State Key Laboratory of Dark Matter Physics, Shanghai Jiao Tong University, China}

\author{Baifei Shen}
\affiliation{Shanghai Normal University, China}
\affiliation{ShanghaiTech University, China}

\author{Liangliang Ji}
\email{jill@siom.ac.cn}
\affiliation{State Key Laboratory of Ultraintense Laser Science and Technology, Shanghai Institute
	of Optics and Fine Mechanics, Chinese Academy of Sciences, China}
\affiliation{ShanghaiTech University, China}

\author{Ruxin Li}
\affiliation{State Key Laboratory of Ultraintense Laser Science and Technology, Shanghai Institute
	of Optics and Fine Mechanics, Chinese Academy of Sciences, China}
\affiliation{ShanghaiTech University, China}

\date{\today}
\begin{abstract}
Paired superradiance (PSR) is a macro-coherent two-photon process capable of very large gain, making it promising for detecting ultra-weak signals induced by neutrinos or dark matter. 
A major goal has been to increase the system volume $V$ and density $n$, since the signal intensity scales as $(nV)^2$.
We recast finite PSR as a parametric amplifier driven by the electromagnetic vacuum.
The usual zero-field semiclassical initial condition is replaced by vacuum inputs fixed by the quantum two-point function.
Combining this formulation with Maxwell--Bloch evolution and finite-length stability analysis,
we find that PSR produces an irreducible vacuum background that can develop into macroscopic bursts once the gain-length product exceeds \(\Gamma L=\pi/2\) for a sufficient coherence time.
These results, together with a closed-form formula for estimating the vacuum-seeded photon yield, 
establish a previously overlooked constraint for high-gain PSR, 
with direct implications for proposed neutrino and dark-matter studies.
\end{abstract}

\maketitle

\textbf{Introduction}.
Macroscopic coherent emission, such as superradiance (SR)~\citep{Dicke:1954zz}, 
dramatically enhances radiative processes by scaling the emission intensity with the square of the participating atom number, $N^2$.
It provides a powerful route for converting weak microscopic transitions into strong optical signals.

In ordinary SR, phase coherence is maintained only over a wavelength-scale distance, 
typically limiting the coherence volume $V_{\rm coh}$ to $\mathcal{O}(\mu\mathrm{m}^3)$.
A two-photon generalization, paired superradiance (PSR), 
overcomes this limitation by emitting photon pairs in opposite directions
~\citep{Yoshimura:2012tm,Miyamoto:2014PTEP:ptu152,Miyamoto:2017JPCA:121.3943,Hiraki:2018jwu,Miyamoto:2015tva}. 
With this phase-matching, $V_{\rm coh}$ can be macroscopically large, 
and $N$  scales with the system size.
Such capability has inspired ambitious proposals to use PSR as a macro-coherent amplifier for ultraweak fundamental processes, 
such as radiative emission of neutrino pairs (RENP)
~\citep{Fukumi:2012rn,Dinh:2012qb,Yoshimura:2013wva,Yoshimura:2015fna,Tanaka:2017juo,Tashiro:2013vja,BoyeroGarcia:2015dye,Vaquero:2016ovj,Song:2015xaa,Zhang:2016lqp,Huang:2019phr,Ge:2021lur,Ge:2022cib,Ge:2023oag} 
and the detection of ultralight dark matter~\citep{Bhoonah:2019eyo,Huang:2019rmc,Sasao:2017zdn}, 
where the massive $N^2=(nV_{\rm coh})^2$ enhancement is crucial.

However, a fundamental quantum constraint has been overlooked.
The prepared macro-coherence makes PSR an amplifier for phase-matched fields,
and any linear amplifier inevitably amplifies electromagnetic vacuum fluctuations~\cite{Caves:1982zz,Caves:2012kxf}. 
Existing theoretical analyses are largely based on semiclassical approaches in which vacuum fluctuations vanish by construction,
and all fields remain zero without an external trigger.
This omission is harmless for existing proof-of-principle experiments
~\citep{Miyamoto:2014PTEP:ptu152,Hiraki:2018jwu,Miyamoto:2015tva,Miyamoto:2017JPCA:121.3943},
where the amplification is moderate.
In order to achieve the extreme amplification required for fundamental physics, however,
the PSR medium must operate in the high-density, large-volume regime,
where the vacuum effects become non-trivial.

In this work, we demonstrate that vacuum fluctuations have significant impacts when PSR is used for probing ultraweak signals.
By explicitly representing vacuum as stochastic fields, 
we obtain the irreducible vacuum background of the PSR output.
We discover a finite-length absolute instability.
Once the density-length product exceeds a critical threshold,
the PSR system becomes unstable against vacuum fluctuations. 
We further provide a closed-form formula for estimating the vacuum-seeded photon yield,
establishing an upper limit for PSR amplification.
These findings establish vacuum-seeded start-up as a fundamental constraint for PSR-based ultraweak signal detection,
and necessitate a careful re-evaluation of proposals whose parameters enter the unstable domain. 
This mechanism is distinct from the QED backgrounds arising from radiative processes of the medium itself~\cite{Yoshimura:2015fna,Tanaka:2017juo}.

\begin{figure}[t]
\begin{center}
\includegraphics[width=0.45\textwidth]{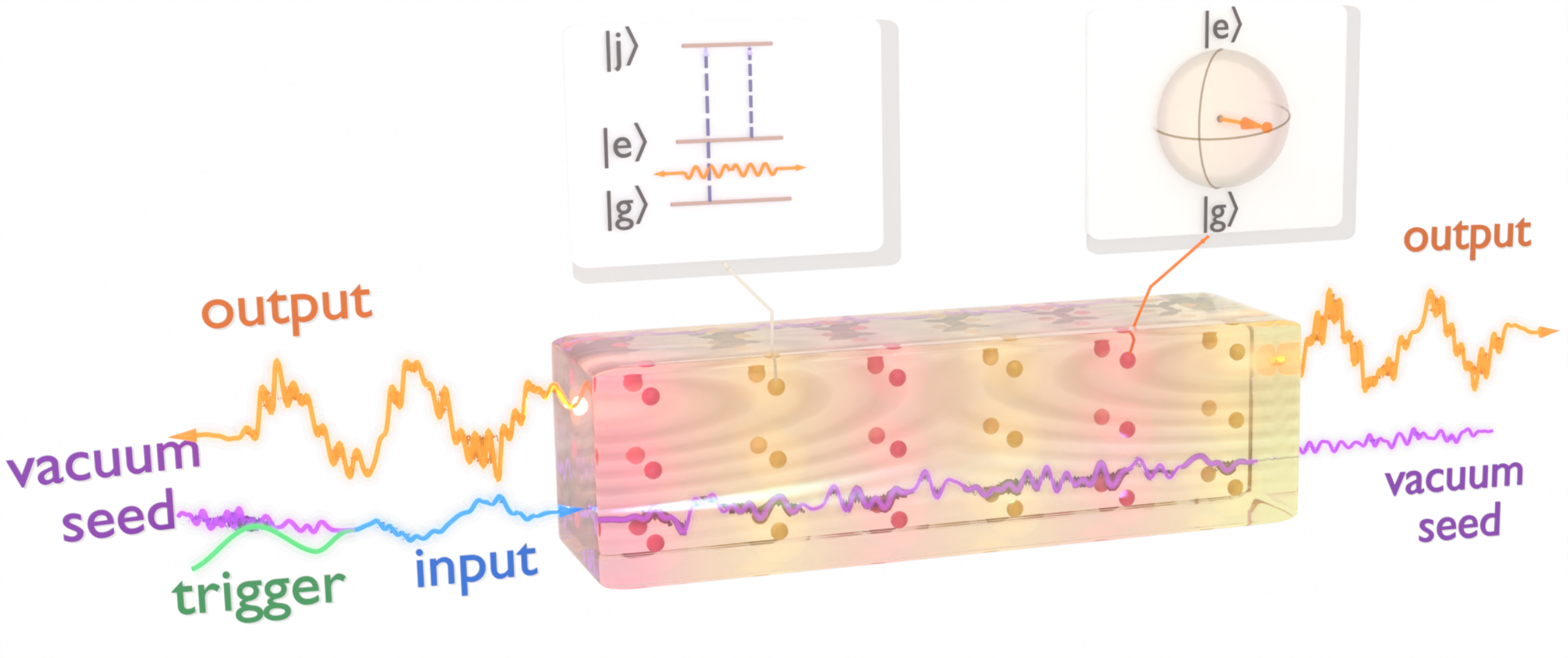}
\end{center}
\vspace{-6mm}
\caption{\label{fig:amplified_noise}
Schematic of paired superradiance in a macro-coherent medium.
A prepared \(|e\rangle\)-\(|g\rangle\) coherence couples counter-propagating phase-matched two-photon modes. 
The system may be seeded either by an injected trigger or by vacuum fluctuations.
The left inset shows the effective two-photon transition via virtual intermediate states,
and the right inset gives a Bloch-sphere representation of the prepared macroscopic coherence.
}
\vspace{-6mm}
\end{figure}

The PSR setup is illustrated in Fig.~\ref{fig:amplified_noise}.
A finite open medium of para-hydrogen ($\mathrm{pH}_2$) molecules are prepared with macro-coherence between the ground state $|g\rangle$ and the lowest vibrationally excited state $|e\rangle$,
separated by an energy gap $\omega_{eg}$.
The $|e\rangle\rightarrow|g\rangle$ transition is dipole-forbidden and proceeds only through far-detuned virtual  states $|j\rangle$ by simultaneous emission of two photons of energy $\Omega=\omega_{eg}/2$ in opposite directions.
This counter-propagating pair structure couples the right- and left-going modes at frequency $\Omega$. 
A right-going field can generate a left-going field, and vice versa.
A trigger field at frequency $\Omega$ entering from either boundary stimulates pair emission from the entire coherent medium.
In the weak trigger limit, the emitted field grows proportionally to the trigger. 
The medium operates as a parametric amplifier, generating large output on both sides.
The same amplification also applies to vacuum fluctuations,
which are present within the medium and continuously enter from the open boundaries.

To describe this dynamics quantitatively, 
we decompose the electric field $E$ into counter-propagating, slowly varying envelopes $\mathcal{E}_{R,L}$ around the carrier frequency $\Omega$,
\[
	E(x,t)=\frac{1}{2}\left[\mathcal{E}_{R}(x,t)\e^{\i\Omega x}+\mathcal{E}_{L}(x,t)\e^{-\i\Omega x}\right]\e^{-\i\Omega t}+\text{c.c.}
\]
We use \(\hbar=c=\epsilon_0=1\), 
and the cycle-averaged flux of one traveling component is \(|\mathcal E_{R,L}|^2/2\).
Far-detuned components are neglected within the slowly varying envelope approximation.
The effective two-photon response of the medium is obtained by applying the Markov approximation and eliminating the virtual states.  
Under these approximations, the field envelopes satisfy
\begin{align}
\left(\partial_t \pm \partial_x\right) \mathcal{E}_{R,L} = -\i \left[ A \mathcal{E}_{R,L} + B \mathcal{E}_{L,R}^* \right],\label{eq:Maxwell_Eqs}\\
A = \frac{\Omega n}{2}\left[\alpha_{gg}\rho_{gg} + \alpha_{ee}\rho_{ee}\right], \qquad B = \frac{\Omega n}{2}\alpha_{eg}\rho_{eg},
\end{align}
where $n$ is the molecular number density.
Here $A$ describes the refractive phase shift and $B$ the coherence-induced pair coupling.
For the effective $|g\rangle$-$|e\rangle$ system,
we denote the density-matrix elements by $\rho_{ab}$ ($a,b\in\{e,g\}$). 
\(\rho_{ee}\) and \(\rho_{gg}\) are the excited- and ground-state populations, 
and \(\rho_{eg}\) is the macroscopic coherence.
The coefficients \(\alpha_{gg}\), \(\alpha_{ee}\), and \(\alpha_{eg}\) are the
effective two-photon polarizabilities of the molecule.
Their microscopic sum-over-state definitions, 
together with the full Maxwell--Bloch equations used in the nonlinear simulations, are given in Appendix~\ref{app:MB_model}.
A detailed derivation and the numerical parameters for $\mathrm{pH}_2$ can be found in Refs.~\cite{Fukumi:2012rn,Bhoonah:2019eyo,Yoshimura:2012tm}.

In the early stage relevant for vacuum-triggered startup, 
the generated field is still weak and does not significantly deplete the prepared medium.  
The density-matrix elements $\rho_{ab}$ can therefore be assumed to evolve independently of the field.  
In particular, the prepared macroscopic coherence follows its natural decay with characteristic decoherence time $T_2$
\begin{equation} \label{eq:rhoeg_decay_main}
\rho_{eg}(t)\simeq \rho_{eg}(0)e^{-t/T_2},
\end{equation}
up to an overall phase that can be absorbed into the definition of the field envelopes 
(see Appendix~\ref{app:MB_model} for details).  

The coefficient $B$ couples the left- and right-going fields.
A mode in one direction generates its counter-propagating partner,
which in turn feeds back into and amplifies the original mode.
The coefficient $|B|$ determines the amplification rate, and we define the instantaneous gain
\begin{equation} \label{eq:Gamma_decay_main}
\Gamma(t)\equiv |B(t)| = \Gamma_0 e^{-t/T_2}, \qquad \Gamma_0= \frac{\Omega n}{2} |\alpha_{eg}\rho_{eg}(0)|.
\end{equation}
The populations \(\rho_{ee}\) and \(\rho_{gg}\) are approximately constant during
this linear startup stage.  

For the analytic stability threshold below we first freeze $A$ and $B$ at their initial values. 
The closed-form formula later includes the exponential decay in Eq.~\eqref{eq:Gamma_decay_main}, 
while the time-domain simulations evolve the full Maxwell--Bloch system.

To analyze the onset of vacuum-triggered PSR, 
we fix $A$ and $B$ as constants,  
and expand the fields as
\begin{equation}
	\mathcal{E}_{R,L}(x,t) = \int \d\kappa  \tilde{\mathcal{E}}_{R,L}(\kappa,t)\e^{-\i\kappa x}.
\end{equation}
The mode vector
$\mathbf{X}_{\kappa} = [\tilde{\mathcal{E}}_{R}(\kappa),\tilde{\mathcal{E}}_{L}^{*}(-\kappa)]^{T}$
obeys
\begin{equation}
	\partial_t \mathbf{X}_{\kappa}=M_{\kappa}\mathbf{X}_{\kappa}, \qquad
	M_{\kappa} = \i\begin{pmatrix}
		\kappa-A & -B         \\
		 B^{*}   & -(\kappa- A)
	\end{pmatrix},
\end{equation}
with eigenvalues
\begin{equation}
	\lambda_{\pm}(\kappa) = \pm\sqrt{|B|^{2}-(\kappa- A)^{2}} .
\end{equation}
Instability occurs for \( |B|>|\kappa-A| \), yielding a band of unstable spatial modes centered at \(\kappa_{0}=A\).
The maximal growth rate is
\begin{equation}\label{eq:Gamma_max}
	\Gamma = | B| = \frac{\Omega n}{2}\,|\alpha_{eg}\rho_{eg}|.
\end{equation}
This analysis indicates that any electromagnetic seed in this phase-matched band is exponentially amplified.

For a finite open medium, local gain must compete with radiative escape through the boundaries.
In order to analyze the stability of the equations \eqref{eq:Maxwell_Eqs},
we impose homogeneous boundary conditions
\(\mathcal E_R(0,t)=0\) and \(\mathcal E_L(L,t)=0\),
and seek growing modes \(\mathcal E_R=r(x)e^{\lambda t}\) and \(\mathcal E_L^*=p(x)e^{\lambda t}\).
Substituting these into Eq.~\eqref{eq:Maxwell_Eqs} gives
\begin{equation}
q\cot(qL)=-\lambda,\qquad \text{where } q^2=\Gamma^2-\lambda^2.
\end{equation}
At threshold we have $\lambda=0$, $qL=\pi/2$ and
\begin{equation}
	\Gamma L=\frac{\pi}{2}.
\end{equation}
Hence \(\Gamma L>\pi/2\) is the condition for absolute instability.
This threshold is mathematically the backward-wave or mirrorless parametric-oscillator threshold~\cite{Harris:1966ApPhL:9.114,Ding:1996IJQE:32.1574,Canalias:2007NaPho:1.459}, with \(\Gamma\) generated here by PSR coherence.
Below threshold, amplification leaks out through the open boundaries before self-sustained growth develops.
The subsequent growth is eventually limited by decoherence and nonlinear depletion of the stored excitation (Appendix~\ref{sec:hybrid_composite_random_seed}).

\textbf{Stochastic Vacuum Input}.
In ordinary one-photon super-fluorescence, 
the quantum seed is commonly assigned to the density-matrix elements~\citep{Polder:1979PhRvA:19.1192}.
In PSR, however, the medium is prepared with a macroscopic two-photon coherence $\rho_{eg}\neq0$.
Consequently, the density-matrix elements are dominated by their mean values and may be treated deterministically at the beginning.
Within the reduced Maxwell--Bloch equations, 
they enter the field dynamics only through the coefficients $A$ and $B$.
Their fluctuations modulate the gain but do not generate the field from zero.

The relevant seeds are therefore the zero-point fluctuations of the electromagnetic field.
These vacuum seeds exist throughout the medium at the initial time and are continuously supplied through the boundaries during the evolution.
This setup corresponds to the windowed gas-target geometry of Refs.~\cite{Miyamoto:2014PTEP:ptu152,Miyamoto:2015tva}.
The vacuum fluctuation is Gaussian and is fully characterized by its two-point function.
We represent it by a stochastic boundary field with the same symmetric-order correlator.
The justification is given in Appendix~\ref{app:stochastic_vacuum_justification}.
Each initial random field is then evolved with the deterministic Maxwell--Bloch equations.

For the numerical simulations, 
we truncate the vacuum spectrum to a bandwidth \(2W\) around the carrier \(\Omega=\omega_{eg}/2\), with \(W\gg\Gamma\).
In the white-noise limit, the correlator is proportional to a \(\delta(t-t')\) kernel.
For finite bandwidth, it is replaced by the corresponding sinc kernel,
\begin{equation} \label{eq:vacuum_corr}
\mathbb E\left[\mathcal E_{\mathrm{in}}(t)\mathcal E_{\mathrm{in}}^*(t^\prime)\right] = \frac{\Omega W}{\pi S}\mathrm{sinc}[W(t-t^\prime)],
\end{equation}
where $\mathrm{sinc}(x)=\sin(x)/x$.

We generate \(\mathcal E_{\rm in}(t)\) numerically as a complex Gaussian random process with this correlator. 
The effective transverse area is fixed by the Fresnel-number condition \(\mathcal F=\Omega S/(2\pi L)=1\)~\citep{Polder:1979PhRvA:19.1192}.

We perform a matched pair of simulations with the same boundary seed. 
One is a reference run \(\mathcal E_{\mathrm{ref}}^{(i)}(t)\), with PSR coupling switched off (\(\alpha_{eg}=0\)), 
and the other is a production run \(\mathcal E_{\mathrm{tot}}^{(i)}(t)\) with the full coupling retained.
The output photon number is inferred from
\begin{equation} \label{eq:intensity_subtraction}
N_{\mathrm{sub}}^{(i)} =\frac{S}{2\Omega}\int \left(|\mathcal E_{\mathrm{tot}}^{(i)}(t)|^{2}-|\mathcal E_{\mathrm{ref}}^{(i)}(t)|^{2}\right)\d t.
\end{equation}
The subtraction removes the universal half-quantum offset mode by mode.
The quantity $N_{\mathrm{sub}}^{(i)}$ is only an estimator, 
whose average and variance  are directly related to the physical photon-number statistics,
$\langle \hat N\rangle=\langle N_{\mathrm{sub}}\rangle$  and $\mathrm{Var}[\hat N]=\mathrm{Var}[N_{\mathrm{sub}}]+\langle N_{\mathrm{sub}}\rangle/2$.
A detailed proof can be found in Appendix~\ref{app:stochastic_vacuum_justification}.
In the linear regime, the matched-subtraction prescription reproduces the quantum result.
Beyond that regime, exact quantum equivalence is no longer claimed.
When nonlinear backreaction sets in, the field is already macroscopically occupied. 
The subtracted vacuum offset is then negligible and the crossover to the semiclassical description is natural.

\textbf{Vacuum-Triggered PSR}.
To quantify the role of vacuum fluctuations, 
we numerically evolve the \(1+1\)D Maxwell--Bloch equations with stochastic vacuum field whose statistics is given by Eq.~\eqref{eq:vacuum_corr}.
In the following we focus on the no-trigger limit of Fig.~\ref{fig:amplified_noise}, in which the output is generated solely from the incident vacuum.
All simulations assume the initially maximally coherent state $\rho_{gg}=\rho_{ee}=\rho_{eg}=1/2$.
The vacuum-seeded output depends on the initial gain \(\Gamma_0L\), the coherence time \(T_2\), 
and the random seed for the vacuum.

\begin{figure}[!htp]
\begin{center}
\includegraphics[width=0.48\textwidth]{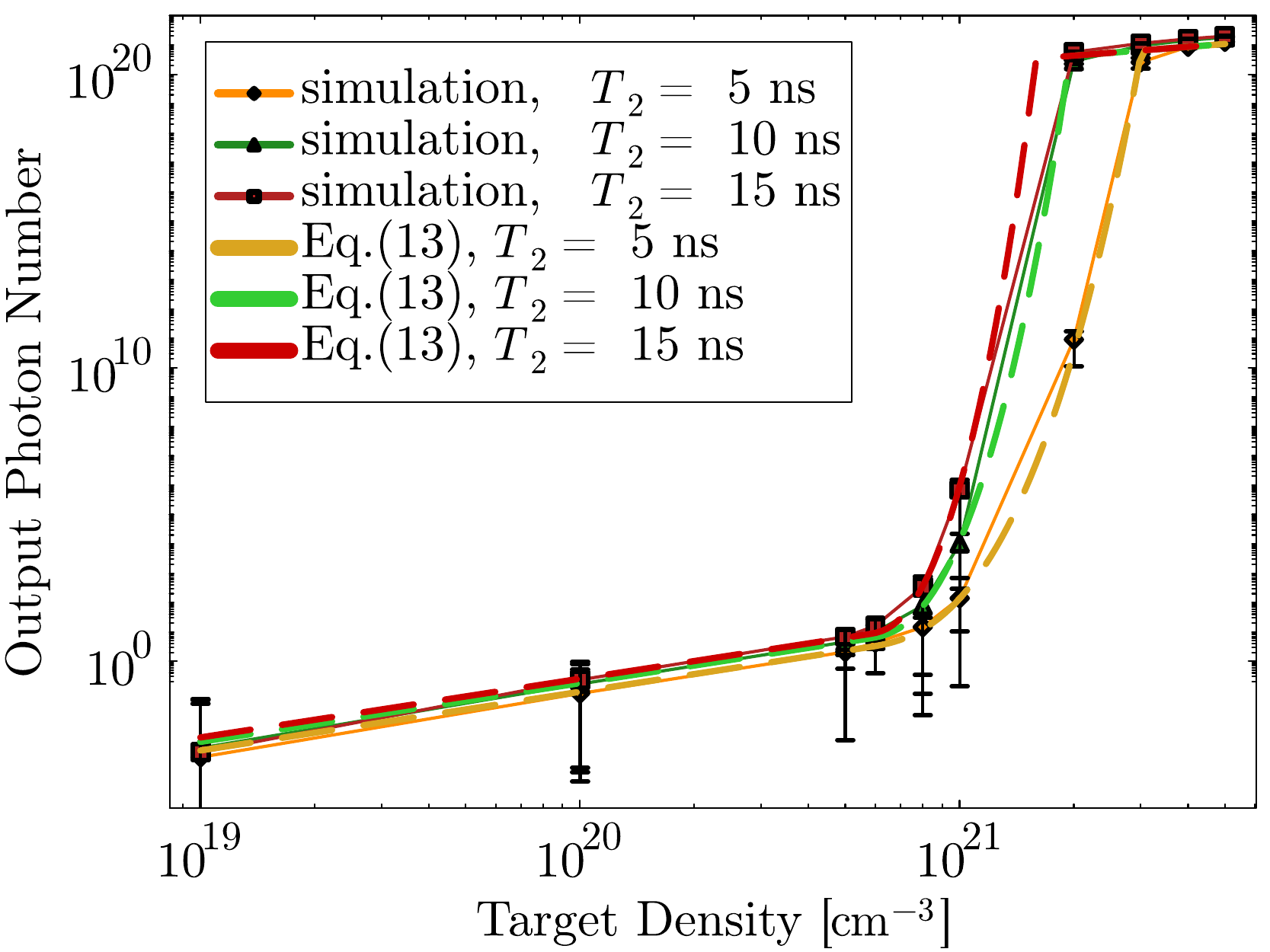}
\end{center}
\vspace{-6mm}
\caption{\label{fig:vacuum_background}
Output photon number versus molecular density for vacuum-seeded runs without a classical trigger.
Target length is fixed at $L=30\,\mathrm{cm}$.
Symbols with error bars show stochastic Maxwell--Bloch simulations. 
The error bars denote one standard deviation of the physical output photon number inferred from the matched-subtraction variance relation, 
and the dashed curves show the corresponding closed-form formula of Eq.~\eqref{eq:quick_selfcontained_main}.}
\end{figure}

Fig.~\ref{fig:vacuum_background} shows the ensemble-averaged output photon number as a function of density for $T_2=5,\,10$ and $15\,\mathrm{ns}$, 
at a fixed medium length $L=30\,\mathrm{cm}$.
The error bars indicate one standard deviation of the physical output photon number, 
obtained from the matched-subtraction variance relation given above.
In the time-dependent case, we denote the initial gain by \(\Gamma_0\), 
with \(\Gamma_0L=(\Omega/2)|\alpha_{eg}\rho_{eg}(0)|\,nL\).
The threshold is common to all three cases and is set by \(\Gamma_0L=\pi/2\).
For \(\Gamma_0L<\pi/2\), the output remains finite and \(\langle N\rangle\propto n^2\).
Above the threshold, the output rises rapidly and, for sufficiently large \(T_2\), can enter the burstlike regime. 
With larger \(T_2\), the amplification persists longer before decoherence, and the post-threshold growth steepens.
At still higher density, nonlinear depletion and back-reaction quickly drive the system toward saturation.

\begin{figure}[!htp]
\begin{center}
\includegraphics[width=0.48\textwidth]{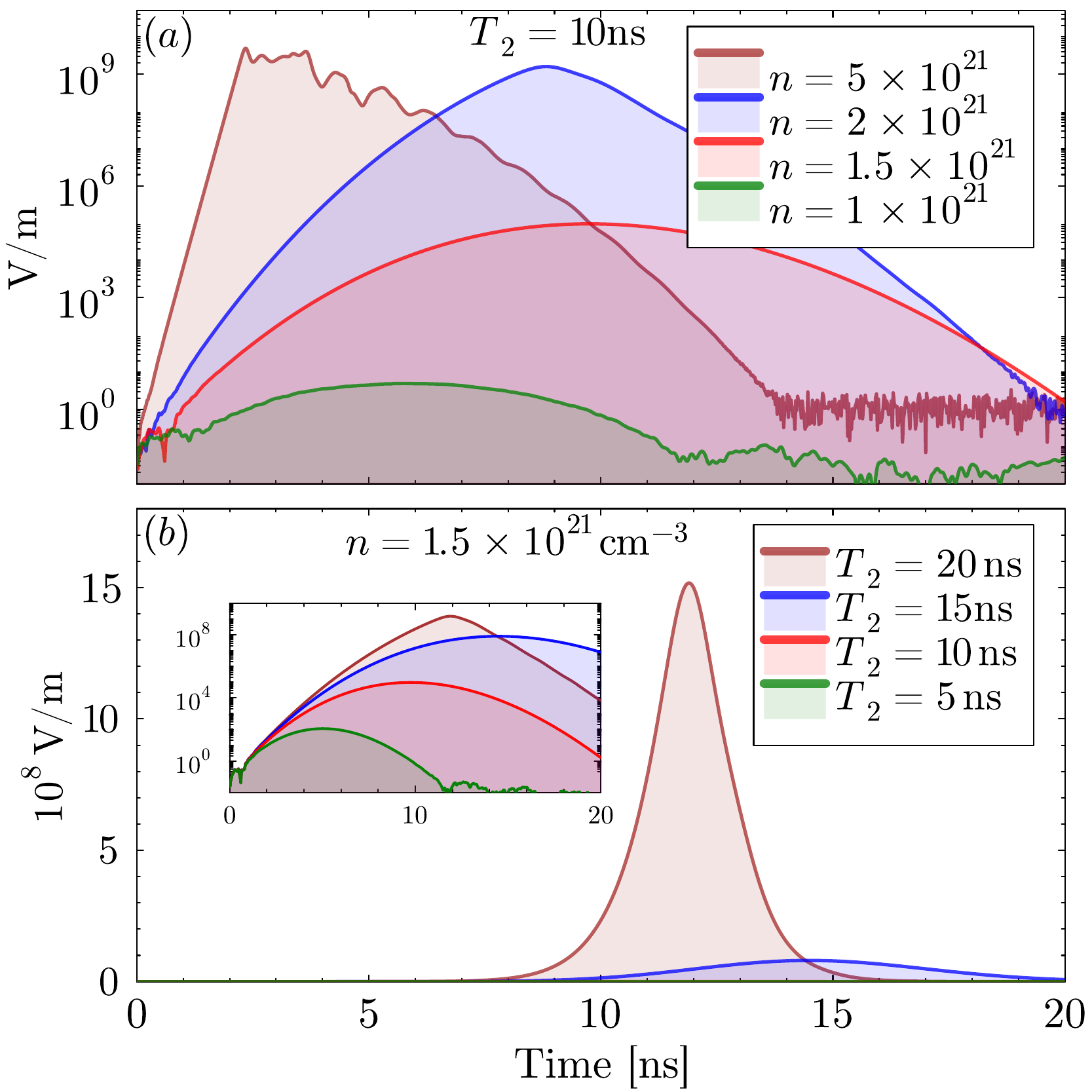}
\end{center}
\vspace{-6mm}
\caption{\label{fig:burst_time}
Simulated field amplitudes of the right-going mode at the exit for different densities and coherence times.
All simulations use the same random seed and a fixed target length \(L=30\,\mathrm{cm}\).
}
\end{figure}

The corresponding field amplitude evolution is shown in Fig.~\ref{fig:burst_time},
where all curves are generated from the same random seed.
The upper panel shows that, at fixed decoherence time, increasing the density accelerates the amplification.
Once saturation is reached, increasing the density further mainly advances the onset of the burst.
The lower panel shows that, at fixed density, the amplification lasts longer for larger coherence times,
allowing the output to reach saturation before decoherence suppresses the gain.
The curves with \(T_2\le 10\,\mathrm{ns}\) remain too small to be resolved on linear scale.

For rapid parameter estimates, we construct a semi-analytic closed-form estimate for the vacuum-seeded photon yield:
\begin{equation} \label{eq:quick_selfcontained_main}
\langle N\rangle \simeq N_\ast(\Gamma_0L)^2 \e^{\left[ \Theta\left(\Gamma_0L-\frac{\pi}{2}\right)
\,\mathcal{S}\left(\Gamma_0L,\frac{T_2}{L}\right)
\right]},
\end{equation}
where $N_\ast$ is the low-gain vacuum normalization, 
and the factor $\mathcal S$ describes the post-threshold growth together with the finite build-up time of the outgoing field.
The approximation scheme and the explicit form of $N_\ast$ and $\mathcal S$ are given in Appendix~\ref{sec:hybrid_composite_random_seed}, Eqs.~\eqref{eq:quick_Phi_final}--\eqref{eq:quick_Nstar_final}.
Eq.~\eqref{eq:quick_selfcontained_main} reproduces the $\langle N\rangle\propto n^2$ scaling in the low-gain regime, 
the $\Gamma_0L=\pi/2$ threshold, 
and the $\ln\langle N\rangle \propto n$ scaling in the large-gain regime.
As the dashed lines in Fig.~\ref{fig:vacuum_background} show, 
the estimate correctly tracks the numerical trend from the low-gain regime through the rapid post-threshold growth, 
until the output is cut off by saturation at \(N_{\rm sat}\sim \rho_{ee}(0)nSL\).

\begin{figure}[!htp]
\begin{center}
\includegraphics[width=0.48\textwidth]{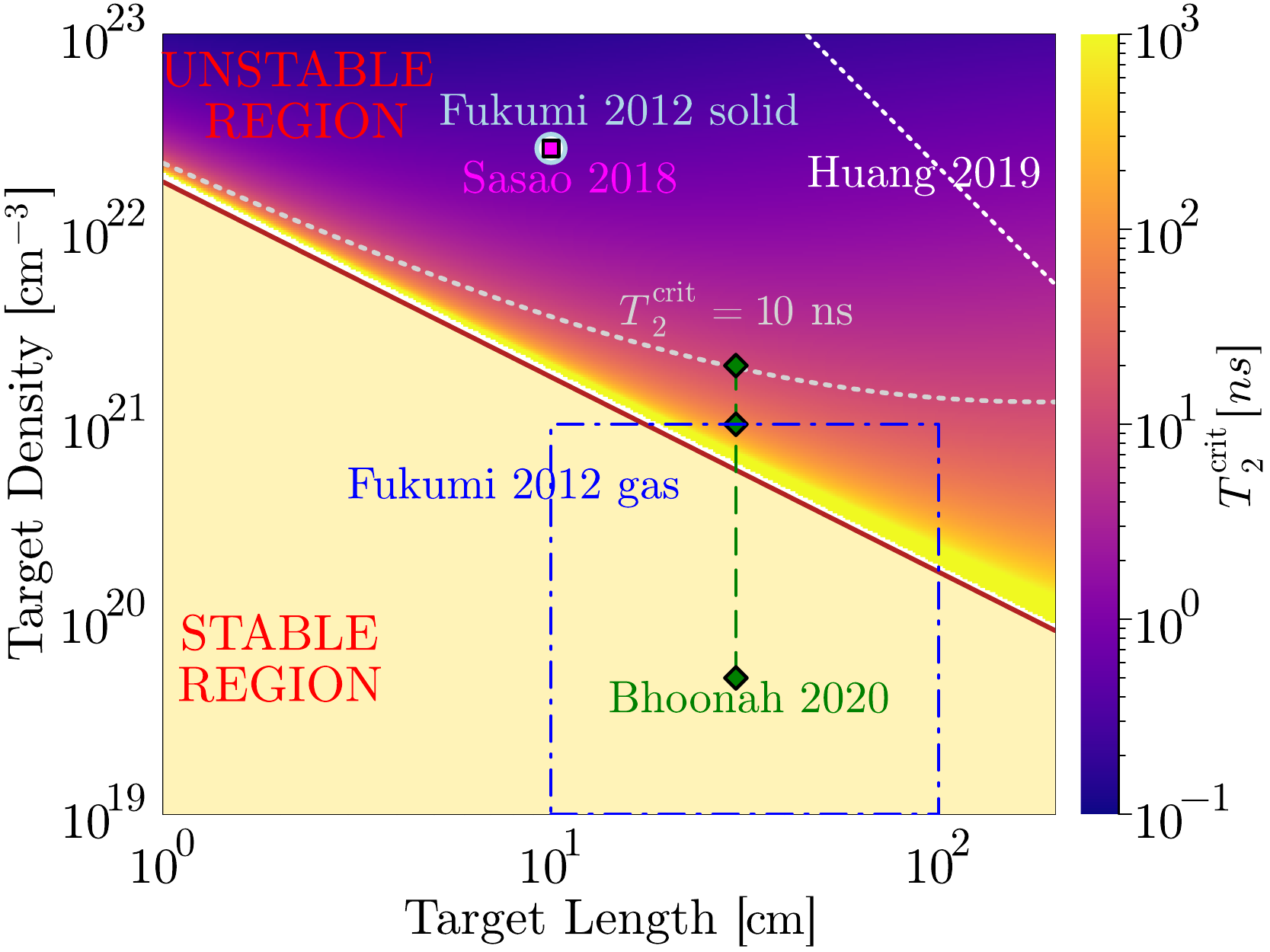}
\end{center}
\vspace{-6mm}
\caption{\label{fig:critical_decoherence_time}
Critical decoherence-time map in the $(L,n)$ plane for the pH\(_2\) benchmark.
Initial maximum coherence $\rho_{gg}=\rho_{ee}=\rho_{eg}=1/2$ is assumed.
The purple solid line is the instability threshold $\Gamma_0L=\pi/2$. 
In the unstable region, the color scale shows the critical coherence time $T_2^{\rm crit}$ for reaching saturation.
Representative parameter ranges quoted for RENP and dark-matter proposals in Refs.~\cite{Bhoonah:2019eyo,Fukumi:2012rn,Huang:2019rmc,Sasao:2017zdn} are shown for comparison.}
\end{figure}

\textbf{Critical Decoherence Time}.
Fig.~\ref{fig:critical_decoherence_time} maps the critical decoherence time required for the unstable PSR channel to reach saturation in the $(L,n)$ plane.
The purple line marks the $\Gamma_0L=\pi/2$ instability threshold.
Below it the medium remains stable, while above it the color scale gives $T_2^{\rm crit}$, defined by requiring Eq.~\eqref{eq:quick_selfcontained_main} to reach $N_{\rm sat}\sim \rho_{ee}(0)nSL$.
The dashed gray contour marks $T_2^{\rm crit}=10\,\mathrm{ns}$.

In Fig.~\ref{fig:critical_decoherence_time}, 
we also present several benchmark parameter choices that have been discussed in the PSR and macrocoherence literature, 
including representative RENP and dark-matter proposals.
This comparison indicates which benchmark parameters fall in the unstable region.
For those above the threshold, the color scale gives the coherence time required for the vacuum-seeded channel to reach the depletion scale.
Because these proposals differ in medium, geometry, and coherence preparation, the comparison is intended only at the order-of-magnitude level in the common $(L,n)$ plane.
These proposed high-gain parameter ranges should be distinguished from existing experiments
~\citep{Miyamoto:2014PTEP:ptu152,Hiraki:2018jwu,Miyamoto:2015tva} 
that have established macrocoherent two-photon dynamics with external triggers, 
but do not yet access the extreme regime where vacuum fluctuations are relevant.
The practical importance of the boundary arises mainly when PSR is pushed toward the much larger gain envisaged for weak-signal amplification.
For signal schemes that use the same unstable PSR channel, vacuum-seeded emission provides a direct competing background and can overwhelm an ultraweak signal.
For schemes in which the signal channel is separated geometrically from the unstable PSR channel, the requirement of avoiding vacuum-triggered startup still restricts the macrocoherent gain achievable in practice.

\textbf{Conclusion}.
We have reconsidered PSR as a parametric amplifier with electromagnetic vacuum as input. 
Whenever the medium is prepared with macroscopic coherence, there exists an irreducible vacuum-seeded output.  
This vacuum-seeded emission is governed by a stability threshold at a gain-length product of $\Gamma L = \pi/2$. 
Below this value, the vacuum generates a stochastic background.  
Above the threshold, the system becomes unstable,  rapidly amplifying the vacuum seed into macroscopic photon bursts. 
The closed-form formula in Eq.~\eqref{eq:quick_selfcontained_main} connects these two regimes and tracks the simulations up to the onset of saturation. 
This intrinsic vacuum background establishes a fundamental limit for PSR-based detectors. 
Current proof-of-principle experiments operate below threshold and remain unaffected.
By contrast, proposals for detecting dark matter or neutrino pairs usually require extreme amplification and might work beyond threshold.  
In such regimes, PSR can start from vacuum fluctuations alone,  producing a competing background and possibly depleting the prepared coherence before the ultraweak signal is measured. 
This vacuum-started emission should be treated as part of the baseline stability and background estimate for any high-gain PSR design.

\textbf{Acknowledgments}.
This work is supported by the National Natural Science Foundation of China (No. 12388102), 
the Strategic Priority Research Program of the Chinese Academy of Sciences (No. XDB0890000), 
the CAS Project for Young Scientists in Basic Research (No. YSBR060).
This work is also supported by the State Key Laboratory of Dark Matter Physics.
N. Song is supported by the 
National Natural Science Foundation of China (NSFC) under Grant Nos. 12475110, 12347105, 12441504 and 12447101.

\bibliographystyle{apsrev4-2}
\bibliography{Ref-PSR}

\clearpage
\onecolumngrid
\section*{Supplemental Material}
\appendix

\section{Maxwell--Bloch Equation for Semiclassical PSR Description} \label{app:MB_model}

In this Appendix we summarize the semiclassical Maxwell--Bloch equations used in our simulations.  
The main text gives only the linearized field equation needed for the analytic stability analysis.
The full simulations also evolve the molecular populations and coherence, thereby including decoherence, depletion, and nonlinear saturation.
We work in the rotating frame at the two-photon resonance $\Omega=\omega_{eg}/2$.

The molecules are described by an effective two-level density matrix $\hat\rho$ in the subspace spanned by $|g\rangle$ and $|e\rangle$.  
We denote its matrix elements by
\begin{equation}
\rho_{ab}\equiv \langle a|\hat\rho|b\rangle, \qquad a,b\in\{g,e\},
\end{equation}
with $\rho_{gg}=1-\rho_{ee}$.  
After adiabatic elimination of the far-detuned intermediate states, 
the effective two-photon interaction gives the following Bloch equations:
\begin{equation} \label{eq:Bloch_Eqs1}
\begin{split}
\dot{\rho}_{ee} &= -\i\left( \mathcal{H}_{eg}^{\rm eff}\rho_{ge} - \mathcal{H}_{ge}^{\rm eff}\rho_{eg} \right) - \frac{\rho_{ee}}{T_1}, \\
\dot{\rho}_{ge} &= -\i\left( \mathcal{H}_{gg}^{\rm eff} - \mathcal{H}_{ee}^{\rm eff} \right)\rho_{ge} - \i\mathcal{H}_{ge}^{\rm eff} \left(\rho_{ee}-\rho_{gg}\right) - \frac{\rho_{ge}}{T_2},
\end{split}
\end{equation}
where $T_1$ is the deexcitation time and $T_2$ is the decoherence time.  
We take $T_1\rightarrow\infty$ in this work.

The effective Hamiltonian matrix elements are
\begin{equation}
\begin{split}
\mathcal{H}_{aa}^{\rm eff} &= -\frac{1}{2} \sum_{m=L,R} \alpha_{aa} |\mathcal{E}_m|^2, \qquad a=e,g, \\
\mathcal{H}_{eg}^{\rm eff} &= -\frac{1}{2} \alpha_{eg} \mathcal{E}_R\mathcal{E}_L, 
\qquad
\mathcal{H}_{ge}^{\rm eff} = \mathcal{H}_{eg}^{\rm eff\,*},
\end{split}
\label{eq:Heff_appendix}
\end{equation}
where $\mathcal{E}_{R,L}$ are the slowly varying field envelops used in the maintext,
and $\alpha_{ab}$'s are he effective polarizabilities and are given by the standard sum-over-states expressions
\begin{equation}
\begin{split}
\alpha_{gg} &= \sum_j |d_{jg}|^2 \left( \frac{1}{\Omega+\omega_{jg}} - \frac{1}{\Omega-\omega_{jg}} \right), \\
\alpha_{ee} &= \sum_j |d_{je}|^2 \left( \frac{1}{\Omega+\omega_{je}} - \frac{1}{\Omega-\omega_{je}} \right), \\
\alpha_{eg} &= 2\sum_j \frac{d_{gj}d_{je}}{\Omega+\omega_{je}},
\end{split}
\label{eq:polarizability_appendix}
\end{equation}
where $d_{ab}$'s denote the electric-dipole matrix element between states \(|a\rangle\) and \(|b\rangle\), 
and \(\omega_{ja}\) is the energy difference between \(|j\rangle\) and \(|a\rangle\).  
The summation runs over far-detuned intermediate states \(|j\rangle\).  

The corresponding slowly varying field equations are
\begin{align}
\left(\partial_t \pm \partial_x\right)\mathcal{E}_{R,L} = -i\left[ A\mathcal{E}_{R,L} + B\mathcal{E}_{L,R}^{*} \right], \label{eq:Maxwell_Eqs_appendix} \\
A= \frac{\Omega n}{2} \left( \alpha_{gg}\rho_{gg} + \alpha_{ee}\rho_{ee} \right), 
\qquad
B= \frac{\Omega n}{2} \alpha_{eg}\rho_{eg},
\label{eq:AB_appendix}
\end{align}
where \(n\) is the molecular number density.  
Equation \eqref{eq:Maxwell_Eqs_appendix} is the same envelope equation used in the main text.  
In the linear stability analysis, \(A\) and \(B\) are held fixed at their initial values, 
whereas in the nonlinear simulations they are updated self-consistently through Eq.~\eqref{eq:Bloch_Eqs1}.

Detailed derivations of the effective Hamiltonian and molecular parameters for
\(\mathrm{pH}_2\) can be found in
Refs.~\cite{Fukumi:2012rn,Bhoonah:2019eyo,Yoshimura:2012tm}.

\paragraph*{Numerical implementation.}

For the \(\mathrm{pH}_2\) benchmark we use the effective polarizabilities from Ref.~\citep{Bhoonah:2019eyo}:
\(\alpha_{gg}=0.90\times10^{-24}\,{\rm cm}^3\),
\(\alpha_{ee}=0.87\times10^{-24}\,{\rm cm}^3\), and
\(\alpha_{eg}=0.0275\times10^{-24}\,{\rm cm}^3\).  
The medium is initially prepared with maximum coherence,
\(\rho_{gg}=\rho_{ee}=1/2\) and \(\rho_{ge}=1/2\).  
Population relaxation is neglected on the simulated time scale; 
coherence dephasing is included through the chosen value of \(T_2\).  
No classical trigger is applied in the vacuum-seeded runs.

The stochastic boundary fields are generated as finite Fourier sums over the detuning $\nu=\omega-\Omega$, 
restricted to the vacuum-noise band $|\nu|\le W$ around the carrier $\Omega$, 
where $W$ is the half-bandwidth of this band and we take $W\gg |B|$ to cover all modes that will be amplified.
Independent complex Gaussian coefficients are used for the two incoming directions, 
with normalization fixed by Eq.~\eqref{eq:vacuum_corr}.  
To initialize the vacuum already present inside the medium at $t=0$, 
the same incoming free-field process is first evolved from an earlier time to the initial time with the PSR coupling switched off.  

Each stochastic realization is then evolved with the deterministic Maxwell--Bloch equations (Eqs.~\eqref{eq:Bloch_Eqs1} and \eqref{eq:AB_appendix}), 
treated numerically as an ODE system after spatial discretization.  
For every density and $T_2$ shown in Fig.~\ref{fig:vacuum_background}, we use 100 matched stochastic pairs.  
The signal run evolves the full Maxwell--Bloch equations, 
while the reference run uses the same boundary seed with the PSR off-diagonal coupling switched off ($\alpha_{eg}=0$).  
The photon number is obtained from the matched-subtraction prescription in Eq.~\eqref{eq:intensity_subtraction}.  
The plotted symbols are ensemble means, and the error bars use the linear-Gaussian variance relation derived in Appendix~\ref{app:stochastic_vacuum_justification}.

\section{Justification of the stochastic vacuum-input model \label{app:stochastic_vacuum_justification}}

At the onset of vacuum-triggered PSR, the field is still weak, 
and the $\mathcal{H}_{ab}$ terms in Eqs.~\ref{eq:Bloch_Eqs1} and \eqref{eq:Heff_appendix} are negligible.
The density matrix elements may decay through the $\rho_{ge}/T_2$ term in Eq.~\eqref{eq:Bloch_Eqs1}, but have not yet been modified by the field. 
The field evolution is then linear. 
In this regime we represent the vacuum by stochastic complex Gaussian boundary fields with the same two-point kernel as the vacuum symmetric-order correlator. 
Linear evolution preserves this correspondence, and a matched coupling-off reference subtraction removes the universal half-quantum offset. 
In this appendix we prove that this procedure yields the correct mean and a simple variance relation for the photon-number estimator used in the simulations.

\subsection{Symmetric-order correspondence}

For one bosonic mode, $[\hat a,\hat a^\dagger]=1$, one defines
\begin{equation}
\hat X=\frac{\hat a+\hat a^\dagger}{2},
\qquad
\hat Y=\frac{\hat a-\hat a^\dagger}{2\i},
\qquad
\hat a=\hat X+\i\hat Y .
\end{equation}
In the vacuum one has $\langle \hat a^\dagger \hat a\rangle=0$, and
\begin{equation} \label{eq:vac_quadrature_moments_appendix}
\langle \hat X\rangle=\langle \hat Y\rangle=0,
\qquad
\langle \hat X^2\rangle=\langle \hat Y^2\rangle=\frac14,
\qquad
\frac12\langle \hat X\hat Y+\hat Y\hat X\rangle=0.
\end{equation}
We introduce classical Gaussian variables $X,Y$ with the same expectation and variance as $\hat{X},\hat{Y}$, and set $\alpha=X+\i Y$.
We then have
\begin{equation} \label{eq:alpha_vac_moments_appendix}
\mathbb E[\alpha]=0,
\qquad
\mathbb E[\alpha^2]=0,
\qquad
\mathbb E[|\alpha|^2]=\frac12,
\end{equation}
which corresponds to the expectations of symmetric vacuum operators, with an additional half-quantum:
\begin{equation}
\mathbb E[|\alpha|^2]
\longleftrightarrow
\frac12\langle \hat a^\dagger \hat a+\hat a\hat a^\dagger\rangle = \langle \hat a^\dagger \hat a\rangle+\frac12 .
\end{equation}

In the linearized PSR stage, the density matrix elements are treated as prescribed $c$-number coefficients.
The field equations are therefore linear and each detected mode is related to the input modes by a linear transformation. 
For one mode, we can write
\begin{equation} \label{eq:linear_evolution_a_appendix}
\hat a(t)=u(t)\hat a(t_0)+v(t)\hat a^\dagger(t_0), \qquad |u(t)|^2-|v(t)|^2=1.
\end{equation}
The vacuum expectation of the symmetric number operator is
\begin{equation} \label{eq:quantum_symmetric_expectation_appendix}
\frac12\left\langle \hat a^\dagger(t)\hat a(t)+\hat a(t)\hat a^\dagger(t)\right\rangle = |v(t)|^2+\frac12.
\end{equation}
For the corresponding classical evolution of the same single mode is:
\begin{equation}
\alpha(t)=u(t)\alpha(t_0)+v(t)\alpha^*(t_0),
\end{equation}
and Eq.~\eqref{eq:alpha_vac_moments_appendix} gives
\begin{equation} \label{eq:classical_intensity_expectation_appendix}
\mathbb E[|\alpha(t)|^2] = \frac12\bigl(|u(t)|^2+|v(t)|^2\bigr) = |v(t)|^2+\frac12,
\end{equation}
which has the same form as Eq.~\eqref{eq:quantum_symmetric_expectation_appendix}.
Thus, for single mode, after the same linear evolution, the classical Gaussian ensemble average
reproduces the symmetric-ordered vacuum expectation value.

The above conclusion can be extend to the multimode boundary field. 
To fix the normalization, one may start from a box-normalized forward-propagating field in an auxiliary volume \(V=S L_q\),
\begin{equation}
\hat E^{(+)}(x,t) = \i\sum_k\sqrt{\frac{\omega_k}{2SL_q}} \hat a_k\e^{\i(kx-\omega_k t)},
\qquad
[\hat a_k,\hat a_{k'}^\dagger]=\delta_{kk'} .
\end{equation}
Taking \(L_q\to\infty\) gives the open-channel form
\begin{equation}
\hat E_{\rm in}^{(+)}(t) = \int_0^\infty \d\omega \mathcal E_{\omega}^{(+)}\hat a_{\omega}^{\rm in} \e^{-\i\omega t},
\end{equation}
where the continuum operators satisfy
\begin{equation}\label{eq:continuum_operator_commutator_appendix}
[\hat a_\omega^{\rm in},\hat a_{\omega'}^{{\rm in}\dagger}] = \delta(\omega-\omega').
\end{equation}
The positive-frequency mode factor obeys
\begin{equation}
\left|\mathcal E_\omega^{(+)}\right|^2=\frac{\omega}{4\pi S}.
\end{equation}

Restricting the input to a finite band \([\Omega-W,\Omega+W]\) around \(\Omega=\omega_{eg}/2\), the band relevant for PSR,
we then extract the fast carrier factor \(\e^{-\i\Omega t}\) and define the slowly varying peak-envelope operator by
\begin{equation}\label{eq:peak_positive_envelope}
\hat E_{\rm in}^{(+)}(t)=\frac12\e^{-\i\Omega t}\hat{\mathcal E}_{\rm in}(t),
\end{equation}
so that
\begin{equation} \label{eq:envelope_operator_appendix}
\hat{\mathcal E}_{\rm in}(t)
\equiv
2\int_{-W}^{W} \d\nu \,\mathcal E_{\Omega+\nu}^{(+)}\hat a_{\Omega+\nu}^{\rm in} \e^{-\i\nu t}
\equiv
\int_{-W}^{W} \d\nu \,\mathcal E_{\Omega+\nu}\hat a_{\Omega+\nu}^{\rm in} \e^{-\i\nu t}
\simeq
\mathcal E_\Omega \int_{-W}^{W} \d\nu \,\hat a_{\Omega+\nu}^{\rm in} \e^{-\i\nu t},
\end{equation}
Here the approximation assumes that the band is narrow on the scale over which
the mode normalization varies, so that
\(\mathcal E_{\Omega+\nu}\simeq \mathcal E_\Omega\).  The definition
\(\mathcal E_{\omega}=2\mathcal E_\omega^{(+)}\) follows from the
peak-envelope convention in Eq.~\eqref{eq:peak_positive_envelope}, which is the
same envelope normalization used in the main text.  Therefore
\begin{equation}
\left|\mathcal E_\Omega\right|^2=4\left|\mathcal E_\Omega^{(+)}\right|^2=\frac{\Omega}{\pi S}.
\end{equation}
The corresponding band-limited vacuum kernel is
\begin{equation} \label{eq:quantum_kernel_band_appendix}
\left\langle 0\left|\hat{\mathcal E}_{\rm in}(t)\hat{\mathcal E}_{\rm in}^\dagger(t^{\prime})\right|0\right\rangle
= |\mathcal E_\Omega|^2 (2W)\,\mathrm{sinc}\left[W(t-t')\right].
\end{equation}
If the classical random amplitudes satisfy
\begin{equation} \label{eq:alpha_nu_stats_appendix}
\mathbb E[\alpha^\textrm{in}(\nu)]=0,
\qquad
\mathbb E[\alpha^\textrm{in}(\nu)\alpha^{\textrm{in}*}(\nu')]=\frac12\delta(\nu-\nu'),
\end{equation}
then the corresponding classical boundary field is
\begin{equation}
\mathcal E_{\rm in}(t)= \mathcal E_\Omega\int_{-W}^{W} \d\nu\,\alpha^\textrm{in}(\nu)\e^{-\i\nu t}
\end{equation}
Using Eq.~\eqref{eq:alpha_nu_stats_appendix}, one finds
\begin{equation}
\mathbb E[\mathcal E_{\rm in}(t)\mathcal E_{\rm in}^*(t')] = \frac12|\mathcal E_\Omega|^2 \int_{-W}^{W}\d\nu\, \e^{-\i\nu(t-t')},
\end{equation}
which is exactly one half of Eq.~\eqref{eq:quantum_kernel_band_appendix}, namely,
\begin{equation} \label{eq:symmetric_kernel_continuum_appendix}
\mathbb E[\mathcal E_{\rm in}(t)\mathcal E_{\rm in}^*(t')]
= \frac12\left\langle \hat{\mathcal E}_{\rm in}(t)\hat{\mathcal E}_{\rm in}^\dagger(t') + \hat{\mathcal E}_{\rm in}^\dagger(t')\hat{\mathcal E}_{\rm in}(t) \right\rangle,
\end{equation}
i.e., the input classical field corresponds to the symmetric order of quantum vacuum.

Using \(|\mathcal E_\Omega|^2=\Omega/(\pi S)\), we have the correlator used in the simulations,
\begin{equation} \label{eq:vacuum_corr_appendix}
\mathbb E[\mathcal E_{\rm in}(t)\mathcal E_{\rm in}^*(t')] = \frac{\Omega W}{\pi S}\, \mathrm{sinc}[W(t-t')],
\end{equation}
which is Eq.~\eqref{eq:vacuum_corr} in the main text. 

The input correspondence is fixed by construction. 
Under linear propagation through the medium, the general output relation is
\begin{equation} \label{eq:linear_output_kernel_appendix}
\hat{\mathcal E}_{\rm out}(t) = \int_{-W}^{W} \d\nu \Bigl[ \mathcal U(t,\nu)\hat a_{\Omega+\nu}^{\rm in} + \mathcal V(t,\nu)\hat a_{\Omega+\nu}^{{\rm in}\dagger} \Bigr],
\end{equation}
where $\mathcal{U}(t,\nu)$ and $\mathcal{V}(t,\nu)$ are the linear input--output kernels. 
The transformation is canonical, so it preserves the field commutator.
The corresponding stochastic field is
\begin{equation} \label{eq:classical_output_kernel_appendix}
\mathcal E_{\rm out}(t) = \int_{-W}^{W} \d\nu \Bigl[ \mathcal U(t,\nu)\alpha^\textrm{in}(\nu) + \mathcal V(t,\nu)\alpha^{\textrm{in}*}(\nu) \Bigr].
\end{equation}
Using Eq.~\eqref{eq:alpha_nu_stats_appendix}, one finds
\begin{equation} \label{eq:classical_output_covariance_appendix}
\mathbb E[\mathcal E_{\rm out}(t)\mathcal E_{\rm out}^*(t')] 
= \frac12 \int_{-W}^{W}\d\nu\, \left[ \mathcal U(t,\nu)\mathcal U^*(t',\nu) + \mathcal V(t,\nu)\mathcal V^*(t',\nu) \right].
\end{equation}

On the quantum side, the only nonvanishing vacuum contractions are
\begin{equation}
\langle \hat a_{\Omega+\nu}^{\rm in}\hat a_{\Omega+\nu'}^{{\rm in}\dagger}\rangle = \delta(\nu-\nu'),
\qquad
\langle \hat a_{\Omega+\nu}^{{\rm in}\dagger}\hat a_{\Omega+\nu'}^{\rm in}\rangle = 0,
\end{equation}
which gives
\begin{equation} \label{eq:output_covariance_symmetric_appendix}
\frac12 \left\langle \hat{\mathcal E}_{\rm out}(t)\hat{\mathcal E}_{\rm out}^\dagger(t') + \hat{\mathcal E}_{\rm out}^\dagger(t')\hat{\mathcal E}_{\rm out}(t) \right\rangle
= \frac12 \int_{-W}^{W}\d\nu\, \left[ \mathcal U(t,\nu)\mathcal U^*(t',\nu) + \mathcal V(t,\nu)\mathcal V^*(t',\nu) \right].
\end{equation}
Thus the stochastic output covariance is the symmetric-ordered output two-point function of the quantum vacuum.  

The single-mode calculation gives the basic correspondence, and the band-limited multimode construction fixes the input normalization.  
Since the subsequent input--output map is linear, a classical Gaussian $c$-field initialized with the vacuum symmetric two-point kernel reproduces the corresponding symmetric-ordered output kernel.

\subsection{Photon-number expectation}

The detected photon number is the time integral of the output photon flux,
\begin{equation} \label{eq:Ndet_time_appendix}
\hat N_{\rm det} = \frac{S}{2\Omega} \int \d t \hat{\mathcal E}_{\rm out}^\dagger(t)\hat{\mathcal E}_{\rm out}(t) .
\end{equation}
Combining Eqs.~\eqref{eq:output_covariance_symmetric_appendix} and \eqref{eq:classical_output_covariance_appendix}, 
and then setting \(t'=t\), the stochastic intensity equals the symmetric-ordered quantum intensity:
\begin{equation} \label{eq:equal_time_symmetric_appendix}
\begin{split}
\mathbb E[|\mathcal E_{\rm out}(t)|^2] 
&= \frac12\left\langle \hat{\mathcal E}_{\rm out}(t)\hat{\mathcal E}_{\rm out}^\dagger(t) + \hat{\mathcal E}_{\rm out}^\dagger(t)\hat{\mathcal E}_{\rm out}(t) \right\rangle \\
&= \left\langle \hat{\mathcal E}_{\rm out}^\dagger(t)\hat{\mathcal E}_{\rm out}(t) \right\rangle + \frac12\left\langle [\hat{\mathcal E}_{\rm out}(t),\hat{\mathcal E}_{\rm out}^\dagger(t)] \right\rangle .
\end{split}
\end{equation}
Because the linear input-output map is canonical, 
the equal-time commutator is a $c$-number and is preserved:
\begin{equation} \label{eq:equal_time_commutator_appendix}
[\hat{\mathcal E}_{\rm out}(t),\hat{\mathcal E}_{\rm out}^\dagger(t)]
=
[\hat{\mathcal E}_{\rm in}(t),\hat{\mathcal E}_{\rm in}^\dagger(t)]
=
2W|\mathcal E_\Omega|^2 .
\end{equation}

In the matched reference run, the pair-production term is zero while the remaining linear propagation is unchanged. 
The reference output field therefore contains only the annihilation-operator component,
\begin{equation} \label{eq:reference_output_kernel_appendix}
\hat{\mathcal E}_{\rm ref}(t) = \int_{-W}^{W} \d\nu\, \mathcal U^{(0)}(t,\nu)\hat a_{\Omega+\nu}^{\rm in},
\end{equation}
with the corresponding classical field
\begin{equation} \label{eq:reference_classical_kernel_appendix}
\mathcal E_{\rm ref}(t) = \int_{-W}^{W} \d\nu\, \mathcal U^{(0)}(t,\nu)\alpha^{\rm in}(\nu).
\end{equation}
Because this number-conserving map has no creation-operator component, it
cannot create photons from the input vacuum.  
Equivalently, the normally ordered vacuum kernel remains zero:
\begin{equation} \label{eq:reference_normal_zero_appendix}
\left\langle \hat{\mathcal E}_{\rm ref}^\dagger(t)\hat{\mathcal E}_{\rm ref}(t) \right\rangle = 0.
\end{equation}
Substituting into Eq.~\eqref{eq:equal_time_symmetric_appendix},
we find the classical reference run contains only the universal half-quantum background.
Subtracting the reference run therefore removes the symmetric-order offset:
\begin{equation} \label{eq:pointwise_subtraction_appendix}
\mathbb E[|\mathcal E_{\rm out}(t)|^2] - \mathbb E[|\mathcal E_{\rm ref}(t)|^2] = \left\langle \hat{\mathcal E}_{\rm out}^\dagger(t)\hat{\mathcal E}_{\rm out}(t) \right\rangle .
\end{equation}
Integrating over the full output time gives
\begin{equation} \label{eq:Nphys_subtraction_appendix}
\langle N_\mathrm{sub}\rangle = \frac{S}{2\Omega}\int \d t\, \left( \mathbb E[|\mathcal E_{\rm out}(t)|^2] - \mathbb E[|\mathcal E_{\rm ref}(t)|^2] \right) = \langle \hat N_{\rm det}\rangle .
\end{equation}
Matched intensity subtraction thus removes the universal half-quantum offset and yields the correct mean photon number. 

\subsection{Photon-number variance}

For the variance, we use the structure of the PSR system:
the right-going output couples only to the conjugate of the left-going input, and vice versa.
In the time domain, this corresponds to the following transformation:
\begin{equation} \label{eq:pairwise_io_kernel_appendix}
\hat{\mathcal E}_{R}^{\rm out}(t) = \int \d t'\, \Bigl[ \mathcal U(t,t')\hat{\mathcal E}_{R}^{\rm in}(t') + \mathcal V(t,t')\hat{\mathcal E}_{L}^{{\rm in}\dagger}(t') \Bigr],
\qquad
\hat{\mathcal E}_{L}^{\rm out}(t) = \int \d t'\, \Bigl[ \mathcal U(t,t')\hat{\mathcal E}_{L}^{\rm in}(t') + \mathcal V(t,t')\hat{\mathcal E}_{R}^{{\rm in}\dagger}(t') \Bigr].
\end{equation}

For vacuum input, the map in Eq.~\eqref{eq:pairwise_io_kernel_appendix} has no source term, so the output field has zero mean.  
Its same-direction two-field correlation also vanishes: $\hat{\mathcal E}_{R}^{\rm out}$ contains an $R$-input annihilation part and an $L$-input creation part, 
and neither can contract with a second $\hat{\mathcal E}_{R}^{\rm out}$ in the product vacuum.  
Therefore
\begin{equation} \label{eq:right_output_zero_correlations_appendix}
\left\langle \hat{\mathcal E}_{R}^{\rm out}(t) \right\rangle = 0.
\qquad
\left\langle \hat{\mathcal E}_{R}^{\rm out}(t)\hat{\mathcal E}_{R}^{\rm out}(t') \right\rangle = 0.
\end{equation}

Define the right-going normally ordered correlation kernel
\begin{equation} \label{eq:right_kernel_appendix}
\mathcal{N}_R(t,t') \equiv \frac{S}{2\Omega} \left\langle \hat{\mathcal E}_{R}^{{\rm out}\dagger}(t') \hat{\mathcal E}_{R}^{\rm out}(t) \right\rangle .
\end{equation}

Here $\mathcal{N}_R$ is Hermitian and can be diagonalized in an orthonormal set of temporal modes $\phi_k(t)$:
\begin{equation} \label{eq:right_kernel_eigen_appendix}
\int \d t'\, \mathcal N_R(t,t')\phi_k(t') = n_k\phi_k(t),
\qquad
\mathcal N_R(t,t') = \sum_k n_k\,\phi_k(t)\phi_k^*(t'), \qquad n_k\ge 0.
\end{equation}
This diagonal basis $\phi_{k}$ may also be used to expand the output field itself:
\begin{equation} \label{eq:right_supermode_def_appendix}
\hat b_{R,k} \equiv \sqrt{\frac{S}{2\Omega}} \int \d t\, \phi_k^*(t)\hat{\mathcal E}_{R}^{\rm out}(t),
\qquad
\hat{\mathcal E}_{R}^{\rm out}(t) = \sqrt{\frac{2\Omega}{S}} \sum_k \phi_k(t)\hat b_{R,k}.
\end{equation}
Using Eqs.~\eqref{eq:right_supermode_def_appendix}, \eqref{eq:right_kernel_appendix}, \eqref{eq:right_kernel_eigen_appendix}, and \eqref{eq:right_output_zero_correlations_appendix}, one obtains
\begin{equation} \label{eq:right_supermode_stats_appendix}
\langle \hat b_{R,k}^\dagger \hat b_{R,\ell}\rangle = n_k\delta_{k\ell},
\qquad
\langle \hat b_{R,k}\hat b_{R,\ell}\rangle = 0.
\end{equation}
Using this basis, the integrated photon number is
\begin{equation} \label{eq:NR_def_appendix}
\hat N_R = \frac{S}{2\Omega} \int \d t\, \hat{\mathcal E}_{R}^{{\rm out}\dagger}(t)\hat{\mathcal E}_{R}^{\rm out}(t) = \sum_k \hat b_{R,k}^\dagger \hat b_{R,k},
\end{equation}
i.e., we have expand the photon number operator as a summation over a set of independent zero-mean operators.
Its expectation and variance can then simply be expressed as
\begin{equation} \label{eq:right_quantum_stats_appendix}
\langle \hat N_R\rangle = \sum_k n_k = \mathrm{Tr}\mathcal N_R,
\qquad
\mathrm{Var}(\hat N_R) = \sum_k (n_k^2+n_k) = \mathrm{Tr}(\mathcal N_R^2)+\mathrm{Tr}\,\mathcal N_R ,
\end{equation}
where
\begin{equation}
\mathrm{Tr}\mathcal N_R = \int \d t \mathcal N_R(t,t),
\qquad
\mathrm{Tr}(\mathcal N_R^2) = \int \d t \int \d t'\, \mathcal N_R(t,t')\mathcal N_R(t',t).
\end{equation}

Next, we derive the variance for the stochastic $c$-field. 
The output fields have the same form as in Eq.~\eqref{eq:pairwise_io_kernel_appendix}
\begin{equation} \label{eq:classical_pair_io_kernel_appendix}
\mathcal E_{R}^{\rm out}(t) = \int \d t' \Bigl[ \mathcal U(t,t')\mathcal E_{R}^{\rm in}(t') + \mathcal V(t,t')\mathcal E_{L}^{{\rm in}*}(t') \Bigr],
\qquad
\mathcal E_{L}^{\rm out}(t) = \int \d t'\, \Bigl[ \mathcal U(t,t')\mathcal E_{L}^{\rm in}(t') + \mathcal V(t,t')\mathcal E_{R}^{{\rm in}*}(t') \Bigr].
\end{equation}
Projecting them onto the same diagonalizing basis as in \eqref{eq:right_kernel_eigen_appendix} and \eqref{eq:right_supermode_def_appendix},
\begin{equation}
\beta_{R,k}^{\rm out} \equiv \sqrt{\frac{S}{2\Omega}} \int \d t\, \phi_k^*(t)\mathcal E_{R}^{\rm out}(t),
\qquad
\beta_{R,k}^{\rm ref} \equiv \sqrt{\frac{S}{2\Omega}} \int \d t\, \phi_k^*(t)\mathcal E_{R}^{\rm ref}(t),
\end{equation}
we have
\begin{equation} \label{eq:beta_covariance_start_appendix}
\mathbb E[\beta_{R,k}^{\rm out}\beta_{R,\ell}^{{\rm out}*}]  = \frac{S}{2\Omega} \int \d t \int \d t'\, \phi_k^*(t)\phi_\ell(t')\, \mathbb E[\mathcal E_{R}^{\rm out}(t)\mathcal E_{R}^{{\rm out}*}(t')] .
\end{equation}

Eqs.~\eqref{eq:classical_output_kernel_appendix} and \eqref{eq:classical_output_covariance_appendix} imply that the classical $c$-number field corresponds to symmetric ordering,
so that
\begin{equation} \label{eq:symmetric_to_normal_kernel_appendix}
\frac{S}{2\Omega}\, \mathbb E[\mathcal E_{R}^{\rm out}(t)\mathcal E_{R}^{{\rm out}*}(t')] = \mathcal N_R(t,t') + \frac12 \mathcal C_R(t,t'),
\end{equation}
where
\begin{equation} \label{eq:commutator_kernel_appendix}
\mathcal C_R(t,t') \equiv \frac{S}{2\Omega} [\hat{\mathcal E}_{R}^{\rm out}(t),\hat{\mathcal E}_{R}^{{\rm out}\dagger}(t')].
\end{equation}

Applying Eq.~\eqref{eq:envelope_operator_appendix} separately to the two input directions and using Eq.~\eqref{eq:continuum_operator_commutator_appendix} gives the band-limited input commutators
\begin{equation} \label{eq:directional_input_commutator_appendix}
[\hat{\mathcal E}_{R,L}^{\rm in}(t),\hat{\mathcal E}_{R,L}^{{\rm in}\dagger}(t')] = |\mathcal E_\Omega|^2 \int_{-W}^{W}\d\nu'\,\e^{-\i\nu'(t-t')},
\end{equation}
The kernels $\mathcal U$ and $\mathcal V$ in Eq.~\eqref{eq:pairwise_io_kernel_appendix} are the Green kernels of the canonical linear field evolution, 
and therefore satisfy the constraint that the field commutator is preserved.  
This is the two-time version of the equal-time statement in Eq.~\eqref{eq:equal_time_commutator_appendix}.  
Substituting Eq.~\eqref{eq:pairwise_io_kernel_appendix} into Eq.~\eqref{eq:commutator_kernel_appendix}, 
using the input commutators \eqref{eq:directional_input_commutator_appendix}, and applying this commutator-preserving constraint gives
\begin{equation} \label{eq:commutator_kernel_explicit_appendix}
\mathcal C_R(t,t') = \frac{S}{2\Omega}[\hat{\mathcal E}_{R}^{\rm out}(t),\hat{\mathcal E}_{R}^{{\rm out}\dagger}(t')] 
= \frac{S}{2\Omega}[\hat{\mathcal E}_{R}^{\rm in}(t),\hat{\mathcal E}_{R}^{{\rm in}\dagger}(t')]
= \frac{S}{2\Omega}|\mathcal E_\Omega|^2 \int_{-W}^{W}\d\nu\,\e^{-\i\nu(t-t')}.
\end{equation}
Using \(|\mathcal E_\Omega|^2=\Omega/(\pi S)\),
Eq.~\eqref{eq:commutator_kernel_explicit_appendix} is the standard projector onto the frequency band \([-W,W]\).  
Since all retained temporal modes lie in this band, it follows that
\begin{equation} \label{eq:commutator_identity_appendix}
\int \d t'\,\mathcal C_R(t,t')\phi_\ell(t') = \phi_\ell(t),
\end{equation}
that is, \(\mathcal C_R\) acts as the identity on the band-limited temporal mode space.

Combining Eqs.~\eqref{eq:beta_covariance_start_appendix}, \eqref{eq:symmetric_to_normal_kernel_appendix}, \eqref{eq:right_kernel_eigen_appendix}, and \eqref{eq:commutator_identity_appendix} then gives 
\begin{equation} \label{eq:beta_out_covariance_diag_appendix}
\mathbb E[\beta_{R,k}^{\rm out}\beta_{R,\ell}^{{\rm out}*}] = \left(n_k+\frac12\right)\delta_{k\ell}.
\end{equation}
For the number-conserving reference run, the normally ordered kernel vanishes
identically, so only the commutator contribution remains.  Let
\(\beta_{L,k}^{\rm ref}\) denote the left-going reference amplitude paired with
\(\beta_{R,k}^{\rm ref}\).  The reference amplitudes in the two directions are
independent vacuum amplitudes,
\begin{equation}
\mathbb E[\beta_{\mu,k}^{\rm ref}\beta_{\eta,\ell}^{{\rm ref}*}]
=
\frac12\delta_{\mu\eta}\delta_{k\ell},
\qquad
\mathbb E[\beta_{\mu,k}^{\rm ref}\beta_{\eta,\ell}^{\rm ref}]
=0,
\qquad
\mu,\eta\in\{R,L\}.
\end{equation}
In particular,
\begin{equation}
\mathbb E[\beta_{R,k}^{\rm ref}\beta_{R,\ell}^{{\rm ref}*}] = \frac12\delta_{k\ell}.
\end{equation}

The coupling-off reference map is number-conserving, 
so it only rotates the input temporal modes and leaves their vacuum statistics unchanged.  
We can therefore use \(\beta_{R,k}^{\rm ref}\) and the paired left-going reference amplitude \(\beta_{L,k}^{\rm ref}\) as the input vacuum amplitudes for the \(k\)-th diagonal channel.  
In that channel the production run may be written as
\begin{equation} \label{eq:single_channel_ref_rep_appendix}
\beta_{R,k}^{\rm out} = u_k\beta^{\rm ref}_{R,k} + v_k\beta_{L,k}^{\mathrm{ref}*},
\end{equation}
where \(u_k\) is the coefficient of the number-conserving propagation part and \(v_k\) is the coefficient of the pair-production part.  
Using the reference-field averages above, 
Eq.~\eqref{eq:single_channel_ref_rep_appendix} gives \(\mathbb E[|\beta_{R,k}^{\rm out}|^2]=(|u_k|^2+|v_k|^2)/2\), 
while Eq.~\eqref{eq:beta_out_covariance_diag_appendix} gives \(\mathbb E[|\beta_{R,k}^{\rm out}|^2]=n_k+1/2\).  
Preservation of the field commutator gives the normalization \(|u_k|^2-|v_k|^2=1\).  
These two relations fix
\begin{equation}
|u_k|^2=1+n_k,\qquad |v_k|^2=n_k.
\end{equation}
The phases of \(u_k\) and \(v_k\) can be absorbed into the definitions of the
reference amplitudes and do not affect the variance below.

The matched-subtraction estimator reads
\begin{equation} \label{eq:NsubR_def_appendix}
N_{{\rm sub},R} = \frac{S}{2\Omega} \int \d t\, \left( |\mathcal E_{R}^{\rm out}(t)|^2 - |\mathcal E_{R}^{\rm ref}(t)|^2 \right)
= \sum_k \left( |\beta_{R,k}^{\rm out}|^2 - |\beta_{R,k}^{\rm ref}|^2 \right).
\end{equation}
For each mode define
\begin{equation}
N_{k,\mathrm{sub}} \equiv |\beta_{R,k}^{\rm out}|^2-|\beta_{R,k}^{\rm ref}|^2,
\end{equation}
so that \(N_{{\rm sub},R}=\sum_k N_{k,\mathrm{sub}}\).  Substituting
Eq.~\eqref{eq:single_channel_ref_rep_appendix} gives
\begin{equation}
N_{k,\mathrm{sub}} = n_k\bigl(|\beta_{R,k}^{\rm ref}|^2+|\beta_{L,k}^{\rm ref}|^2\bigr) + u_kv_k^*\,\beta_{R,k}^{\rm ref}\beta_{L,k}^{\rm ref} + u_k^*v_k\,\beta_{R,k}^{{\rm ref}*}\beta_{L,k}^{{\rm ref}*} .
\end{equation}
Using the Gaussian averages
\begin{equation}
\mathrm{Var}(|\beta_{R,k}^{\rm ref}|^2) = \mathrm{Var}(|\beta_{L,k}^{\rm ref}|^2) = \frac14, 
\qquad 
\mathbb E[|\beta_{R,k}^{\rm ref}\beta_{L,k}^{\rm ref}|^2] = \frac14,
\end{equation}
and the fact that all mixed products with an unmatched Gaussian factor average
to zero, one obtains
\begin{equation}
\mathrm{Var}(N_{k,\mathrm{sub}}) = n_k^2+\frac12 n_k .
\end{equation}

Since different \(k\) are independent, the variance of the right-going
matched-subtraction estimator is
\begin{equation} \label{eq:NsubR_var_total_appendix}
\mathrm{Var}(N_{{\rm sub},R}) = \sum_k\left(n_k^2+\frac12 n_k\right) = \mathrm{Tr}(\mathcal N_R^2) + \frac12 \mathrm{Tr}\,\mathcal N_R.
\end{equation}
Using the right-going version of Eq.~\eqref{eq:Nphys_subtraction_appendix},
which gives
\(\mathbb E[N_{{\rm sub},R}] = \langle \hat N_R\rangle\), one may rewrite the
variance relation as
\begin{equation} \label{eq:variance_relation_appendix}
\mathrm{Var}(\hat N_R) = \mathrm{Var}(N_{{\rm sub},R}) + \frac12 \mathbb E[N_{{\rm sub},R}] .
\end{equation}
Thus the simulated matched-subtraction estimator determines both the mean and
the variance of the quantum vacuum output photon number.

Equations~\eqref{eq:Nphys_subtraction_appendix} and \eqref{eq:variance_relation_appendix} are exact in the linearized PSR regime.
Outside it, the matched-subtraction quantities can still be computed numerically, but the Gaussian identities above no longer apply.

\section{Closed-Form Estimate for Vacuum-Seeded Emission \label{sec:hybrid_composite_random_seed}}

This Appendix derives the closed-form estimate Eq.~\eqref{eq:quick_selfcontained_main} for the vacuum-seeded photon yield.
It provides a fast way to evaluate the vacuum-noise background over a broad range of parameters, 
without running a full time-domain simulation for every parameter set.  
Below threshold the estimate reproduces the quadratic scaling; above threshold it gives exponential amplification.

\subsection{Linearized field problem and stochastic seed}\label{app:threshold}

In the linear regime of the Maxwell--Bloch equations \eqref{eq:Maxwell_Eqs},
$\mathcal{E}_{R}(x,t)$ and $\mathcal{E}_{L}(x,t)$ denote the slowly varying right- and left-moving field envelopes and satisfy
\begin{equation}
	(\partial_t \pm \partial_x)\mathcal{E}_{R,L}= -\i\bigl[A\mathcal{E}_{R,L}+B\mathcal{E}_{L,R}^*\bigr],
\end{equation}
where $A$ and $B$ are given in Eq.~\eqref{eq:Maxwell_Eqs}.
To remove the diagonal phase shift and simplify the off-diagonal coupling, we go to a rotating frame.
Define the rotating-frame fields
\begin{equation} \label{eq:hc_rotating_definition}
	F_{R}(x,t)\equiv \mathcal{E}_{R}(x,t)\,\e^{\i Ax},
	\qquad
	F_{L}(x,t)\equiv \mathcal{E}_{L}(x,t)\,\e^{-\i Ax},
\end{equation}
which absorbs the homogeneous real phase shift \(A\); 
after an additional constant phase rotation of \(F_R\) and \(F_L\), 
the off-diagonal coupling can be written as \(\i\Gamma(t)\) with \(\Gamma(t)=|B(t)|\), giving
\begin{equation} \label{eq:hc_field_eqs}
	(\partial_t+\partial_x)F_R=\i\Gamma(t)F_L^*,
	\qquad
	(\partial_t-\partial_x)F_L=\i\Gamma(t)F_R^*,
\end{equation}
where the effective coupling strength is
\begin{equation} \label{eq:hc_Gamma}
	\Gamma(t)=\Gamma_0 \e^{-t/T_2},
	\qquad
	\Gamma_0=\frac{\Omega n}{2}|\alpha_{eg}\rho_{eg}(0)|.
\end{equation}
Here $x\in[0,L]$ is the longitudinal coordinate and $t$ is time; 
the material parameters entering \(\Gamma_0\) are as defined in the main text.
The exponential form in Eq.~\eqref{eq:hc_Gamma} assumes $\rho_{ge}(t)\simeq\rho_{ge}(0)\e^{-t/T_2}$ throughout the linear stage;
field-induced modification of the coherence is neglected.

The stochastic seed is most simply represented by zero-mean complex Gaussian boundary fields \(\xi_{R,L}\), 
which enter continuously from the two open boundaries:
\begin{align}
	F_R(0,t)=\xi_R(t), &\qquad F_L(L,t)=\xi_L(t), \label{eq:hc_boundary_seed}\\
	F_R(x,0)=\xi_R(-x),& \qquad F_L(x,0)=\xi_L(x-L), \label{eq:hc_initial_seed}
\end{align}
with correlations.
The exact vacuum correlation for a band-limited peak-envelope input of width $2W$ around $\Omega$ is given by Eq.~\eqref{eq:vacuum_corr}:
\begin{equation}
	\mathbb E[\mathcal E_{\rm in}(t)\mathcal E_{\rm in}^*(t')] = \frac{\Omega W}{\pi S}\mathrm{sinc}[W(t-t')].
\end{equation}
In the high-bandwidth limit $W\gg\Gamma_0$ (white-noise approximation),
$\mathrm{sinc}[W(t-t')]\to (\pi/W)\delta(t-t')$. We then use
\begin{equation} \label{eq:hc_correlations}
	\langle\xi_\mu(t)\xi_\nu^*(t')\rangle=\mathcal{B}\,\delta_{\mu\nu}\delta(t-t'),
\end{equation}
where $\mu,\nu\in\{R,L\}$ label the propagation direction.
From the limit of Eq.~\eqref{eq:vacuum_corr}, the noise strength is $\mathcal{B}= \frac{\Omega}{S}$.
This follows from $\frac{\Omega W}{\pi S}\mathrm{sinc}[W(t-t')] \to \frac{\Omega}{S}\delta(t-t')$ when $W\gg\Gamma_0$.

The rotating-frame transformation \eqref{eq:hc_rotating_definition} is a pure phase redefinition
and leaves field intensities invariant.
The right-going output photon number collected over an observation window $T_{\rm obs}$ is therefore the same quantity defined in Eqs.~\eqref{eq:intensity_subtraction} and \eqref{eq:Nphys_subtraction_appendix},
\begin{equation} \label{eq:hc_DeltaN_def}
\langle N_{\rm sub}\rangle= \frac{S}{2\Omega}\int_0^{T_{\rm obs}}\d t
\left(\left\langle \left|F_R(L,t)\right|^2\right\rangle-\left\langle \left|F_R(L,t)\right|^2\right\rangle_{\Gamma=0}\right),
\end{equation}
where $S=2\pi L/\Omega$ is the effective transverse area for Fresnel number $\mathcal{F}=1$.

To characterize the onset of exponential growth, we approximate the gain as constant, 
\(\Gamma(t)\to\Gamma\), and drop the boundary driving.
The correction fields then satisfy homogeneous open-boundary conditions, \(F_R(0,t)=0\) and \(F_L^*(L,t)=0\). 
Seeking normal modes of the linear system \eqref{eq:hc_field_eqs} of the form \(F_R=r(x)e^{\lambda t}\) and \(F_L^*=p(x)e^{\lambda t}\) gives
\((\lambda+\partial_x)r=\i\Gamma p\) and \((\lambda-\partial_x)p=-\i\Gamma r\).
Eliminating \(p\) yields \(r''+q^2 r=0\) with \(q^2\equiv \Gamma^2-\lambda^2\).
Imposing \(r(0)=0\) and \(p(L)=0\) then gives the dispersion relation
\begin{equation} \label{eq:hc_dispersion}
	q\cot(qL)=-\lambda, \qquad q^2+\lambda^2=\Gamma^2.
\end{equation}
At threshold $\lambda=0$, so $qL=\pi/2$ and hence
\begin{equation} \label{eq:hc_threshold}
	\Gamma L=\frac{\pi}{2}.
\end{equation}
Restoring the time dependence of Eq.~\eqref{eq:hc_Gamma}, 
the gain decreases and the system remains above threshold only during \(\Gamma(t)L>\pi/2\).

\subsection{Closed-form yield estimation\label{sec:quick_law_derivation}}

We now build an explicit estimate for \(\langle N_{\rm sub}\rangle\) from three pieces: 
the low-gain normalization Eq.~\eqref{eq:quick_DI_born}, an approximation to the integrated above-threshold exponential amplification, 
and a correction that accounts for the finite time needed for the amplified field to form at the output boundary.

\subsubsection{Low-density normalization}

For weak gain \(\Gamma_0 L\ll1\), we fix the normalization by a Born expansion in the coupling \(\Gamma(t)\) around free propagation. 
We expand the exit field as \(F_R(L,t)=F_R^{(0)}(t)+F_R^{(1)}(t)+F_R^{(2)}(t)+\cdots\), where \(F_R^{(m)}=O(\Gamma^m)\). 
The zeroth-order term is simply the transmitted right-boundary seed,
\(F_R^{(0)}(t)=\xi_R(t-L)\), which is removed by the subtraction in Eq.~\eqref{eq:hc_DeltaN_def}.

At first order, we insert the free left-moving solution \(F_L^{*(0)}(x,t)=\xi_L^*(t-L+x)\) into Eq.~\eqref{eq:hc_field_eqs} 
and integrate along a right-going characteristic, which gives
\begin{equation}
	F_R^{(1)}(t)= \i\int_{\max(0,t-L)}^{t}\d u\,\Gamma(u)\,\xi_L^*(2u-t).
\end{equation}

We define the subtracted intensity
\begin{equation}
	\Delta I(t)\equiv\langle|F_R(L,t)|^2\rangle-\langle|F_R^{(0)}(t)|^2\rangle.
\end{equation}

There is no \(O(\Gamma)\) contribution to $\Delta I(t)$ because \(\langle F_R^{*(0)}(t)F_R^{(1)}(t)\rangle=0\): 
\(F_R^{(0)}\) depends only on \(\xi_R\) while \(F_R^{(1)}\) depends only on the independent seed \(\xi_L\).
Using the white-noise correlator \eqref{eq:hc_correlations}, this gives
\begin{equation}
	\left\langle|F_R^{(1)}(t)|^2\right\rangle = \frac{\mathcal B}{2}\int_{\max(0,t-L)}^{t}\Gamma(u)^2\,\d u .
\end{equation}

The second \(O(\Gamma^2)\) contribution comes from the same boundary seed \(\xi_R\): 
the right-going vacuum can be converted into a left-moving component via the conjugated transport equation \((\partial_t-\partial_x)F_L^*=-\i\Gamma(t)F_R\) 
and then converted back before reaching \(x=L\). 
This enters \(\Delta I(t)\) through
\(2\Re\langle F_R^{*(0)}F_R^{(2)}\rangle=\frac{\mathcal B}{2}\int_{\max(0,t-L)}^{t}\Gamma(u)^2\,\d u\),
where the delta correlation \eqref{eq:hc_correlations} collapses the two-leg time integral in the same way as above,
so the two \(O(\Gamma^2)\) pieces add to
\begin{equation} \label{eq:quick_DI_born}
	\Delta I_{\rm st}^{\rm(LO)}(t) = \mathcal B \int_{\max(0,t-L)}^{t}\Gamma(u)^2\d u .
\end{equation}
Equation~\eqref{eq:quick_DI_born} is the leading stable contribution to the observable \eqref{eq:hc_DeltaN_def} under the same boundary-noise model as used in the simulations.

The corresponding leading-order photon number is therefore
\begin{equation}
	N_{\rm low} \equiv N_{\rm sub}^{\rm(LO)} = \frac{S}{2\Omega}\int_0^{T_{\rm obs}}\Delta I_{\rm st}^{\rm(LO)}(t)\,\d t.
\end{equation}
Since Eq.~\eqref{eq:quick_DI_born} is quadratic in \(\Gamma\), 
and \(\Gamma(u)=\Gamma_0e^{-u/T_2}\), 
this low-density output scales as \((\Gamma_0 L)^2\). 
We therefore define the dimensionless prefactor
\(N_*\) by
\begin{equation}
	N_{\rm low}=N_*(\Gamma_0 L)^2.
\end{equation}
Substituting Eq.~\eqref{eq:quick_DI_born} into the photon-number definition
Eq.~\eqref{eq:hc_DeltaN_def} and factoring out \((\Gamma_0 L)^2\) then gives
\begin{equation} \label{eq:quick_Nstar_born_def}
	N_*= \frac{1}{2(\Gamma_0 L)^2} \int_0^{T_{\rm obs}}dt \int_{\max(0,t-L)}^t \Gamma(u)^2\,du .
\end{equation}
Evaluating the double integral in Eq.~\eqref{eq:quick_Nstar_born_def} gives for \(T_{\rm obs}\ge L\),
\begin{equation} \label{eq:quick_Nstar_exact}
	N_*= \frac{T_2}{8L^2} \left[ 2L - T_2 e^{-2(T_{\rm obs}-L)/T_2} + T_2 e^{-2T_{\rm obs}/T_2} \right].
\end{equation}
In the long-window limit \(T_{\rm obs}\to\infty\),
Eq.~\eqref{eq:quick_Nstar_exact} reduces to the simple uniform form
\begin{equation} \label{eq:quick_Nstar_infty}
	N_*^{(\infty)}= \frac{T_2}{4L}.
\end{equation}

\subsubsection{Integrated unstable action}

Above threshold, the amplification is controlled by the principal unstable branch of the frozen dispersion relation. 
We approximate the growth rate by an elementary function and integrate it over the decaying coupling \(\Gamma(t)\).

For a frozen gain \(\Gamma\), the unstable mode grows as \(e^{\lambda t}\), 
where \(\lambda(\Gamma L)\) is obtained by solving the dispersion relation \eqref{eq:hc_dispersion} for the principal branch. 
We replace the transcendental inversion by an explicit interpolation in the distance to threshold. 
We define
\begin{equation} \label{eq:quickelem_hdelta}
	\delta\equiv \Gamma L-\frac{\pi}{2}, \qquad a\equiv 1-\frac{\pi}{4}.
\end{equation}
We then approximate \(\lambda(\Gamma L)\) by an elementary interpolation that enforces \(\lambda=0\) for \(\delta<0\), 
reproduces the linear threshold behavior \(\lambda L\simeq(\pi/2)\delta\) for \(\delta\to0^+\), 
and approaches \(\lambda\simeq\Gamma\) for \(\Gamma L\gg1\),
\begin{equation}  \label{eq:quick_lambda_eff}
	\lambda_{\rm interp}(\Gamma L)= \Theta(\delta)\, \frac{(\Gamma L)\,\delta(1+a\delta)}{L(1+\delta+a\delta^2)}.
\end{equation}
With \(\Gamma(t)=\Gamma_0\e^{-t/T_2}\), the unstable interval ends when \(\Gamma(t)L=\pi/2\), which gives
\begin{equation} \label{eq:quick_tc}
	t_c=T_2\ln\left(\frac{\Gamma_0 L}{\pi/2}\right)\Theta(\Gamma_0 L-\pi/2).
\end{equation}
Since the instantaneous growth rate \(\lambda(t)\) amplifies the field envelope as \(\mathcal E(t)\simeq \mathcal E(0)\exp[\int_0^t\lambda(t')\d t']\), 
we define \(H\) as the accumulated amplitude-gain exponent during the above-threshold interval. 
Without output-formation delay, the field amplitude would be amplified by \(e^H\), and the photon number by \(e^{2H}\):
\begin{equation} \label{eq:quick_Heff}
	H=\int_0^{t_c}\lambda_{\rm interp}\bigl((\Gamma_0 L)\e^{-t/T_2}\bigr)\d t =\frac{T_2}{L}\Phi(\Gamma_0 L),
\end{equation}
where \(\Phi\) is evaluated in elementary functions. 
Writing \(\Delta\equiv \sqrt{\pi-3}\), one finds
\begin{equation} \label{eq:quick_Phi}
	\Phi(\Gamma_0 L)= \Theta(\delta)\left[ \delta - \frac{1}{\Delta} \ln\!\left( \frac{(2a\delta+1-\Delta)(1+\Delta)} {(2a\delta+1+\Delta)(1-\Delta)} \right) \right],
\end{equation}
where here \(\delta\equiv \Gamma_0 L-\pi/2\).

\subsubsection{Output-formation cumulant}

The factor \(e^{2H}\) would be the photon-number gain for a seed that is already projected onto the unstable spatial mode at the beginning of the above-threshold stage.  
In reality, the vacuum seed enters the medium as freely propagating plane waves and is gradually shaped into the unstable mode through the coupling \(\Gamma(t)\).  
Integrating the first equation in Eq.~\eqref{eq:hc_field_eqs} along a right-going characteristic shows that the right-boundary field at time \(t\) is assembled from contributions generated at earlier times throughout the medium.  
A contribution generated late in the above-threshold interval therefore experiences only a fraction of the full exponent \(H\) before the coupling falls below threshold.

We model this delay by introducing a single effective output-formation time \(\tau_{\rm out}\).
To estimate it, we use the frozen mode at threshold.  
At threshold, \(\lambda=0\) and \(qL=\pi/2\), 
and Eq.~\eqref{eq:hc_dispersion} gives the simple profiles
\begin{equation}
	r(x)\propto \sin\left(\frac{\pi x}{2L}\right), \qquad p(x)\propto -\i\cos\left(\frac{\pi x}{2L}\right).
\end{equation}

The formation of a right-boundary signal can be understood as a two-stage process.  
In the first stage, the right-going field propagates from the left boundary into the medium and is converted to a left-going wave through the coupling.  
In the second stage, the left-going wave is converted back to a right-going wave and propagates to the right boundary.  
The characteristic propagation distance of each stage is estimated by the intensity-weighted mean position of the corresponding threshold-mode profile, 
using the spatial probability densities \(|r(x)|^2\) and \(|p(x)|^2\) respectively.
Since the two stages act in series, the total mean formation time is their sum,
\begin{equation} \label{eq:quick_tau_out}
	\tau_{\rm out} \equiv \frac{\int_0^L x\,|r(x)|^2\,\d x}{\int_0^L |r(x)|^2\,\d x} + \frac{\int_0^L (L-x)\,|p(x)|^2\,\d x}{\int_0^L |p(x)|^2\,\d x} = L\left(1+\frac{4}{\pi^2}\right)\approx 1.405L.
\end{equation}
In our units \(c=1\), \(L\) is the light-crossing time, and \(\tau_{\rm out}\) is fixed once the geometry is specified.

The time available for forming and amplifying the unstable mode is the above-threshold duration itself, so we define
\begin{equation} \label{eq:quickelem_x}
	x_+\equiv \frac{t_c}{\tau_{\rm out}}.
\end{equation}

A contribution generated at a time \(t\) within the above-threshold interval \(0\le t\le t_c\) has an effective ``age'' \(\Delta t\equiv t_c-t\), 
i.e., the time remaining before the coupling drops below threshold.  
A late contribution (\(\Delta t\) small) experiences almost no amplification, 
while an early one (\(\Delta t\gg\tau_{\rm out}\)) experiences essentially the full gain exponent \(H\).  
We model this reduction by a single-timescale response, introducing the dimensionless age \(u\equiv\Delta t/\tau_{\rm out}\) and writing the effective amplitude-gain exponent as
\begin{equation} \label{eq:quick_Heff_delay}
	H_{\rm eff}(u)=H\,f(u), \qquad f(u)\equiv 1-e^{-u}, \qquad 0\le u\le x_+,
\end{equation}
where \(x_+\equiv t_c/\tau_{\rm out}\) sets the maximum dimensionless age.

Because the vacuum seed is supplied uniformly throughout the above-threshold interval, 
the net photon-number enhancement is the uniform average of \(e^{2H_{\rm eff}(u)}\) over \(u\in[0,x_+]\),
\begin{equation}
	\left\langle e^{2Hf}\right\rangle \equiv \frac{1}{x_+}\int_0^{x_+}e^{2Hf(u)}\,\d u .
\end{equation}
We parameterize this average through the first two cumulants of the kernel \(f(u)\).  
Its uniform mean,
\begin{equation} \label{eq:quick_Rout}
	\mu_+(x_+)\equiv \frac{1}{x_+}\int_0^{x_+}f(u)\,\d u = \Theta(x_+)\left[1-\frac{1-e^{-x_+}}{x_+}\right],
\end{equation}
is the average fraction of the full exponent \(H\) experienced across the interval.
The variance of the kernel is
\begin{equation} \label{eq:quick_sigma_def}
	\sigma_+^2(x_+)\equiv \frac{1}{x_+}\int_0^{x_+} f(u)^2\,\d u-\mu_+(x_+)^2,
\end{equation}
which can be evaluated explicitly:
\begin{equation} \label{eq:quick_sigma_explicit}
	\sigma_+^2(x_+)= \Theta(x_+)\left[ 1-\frac{2(1-e^{-x_+})}{x_+} +\frac{1-e^{-2x_+}}{2x_+} -\mu_+(x_+)^2 \right].
\end{equation}

The exact average \(\langle e^{2Hf}\rangle\) for \(f(u)=1-e^{-u}\) is not an elementary function of \(H\) and \(x_+\) and involves exponential-integral functions. 
We therefore use a minimal interpolation for its logarithm.  
Writing \(z=2H\), the small-\(z\) cumulant expansion is
\begin{equation}
	\ln\langle e^{zf}\rangle = z\mu_+(x_+)+\frac{z^2}{2}\sigma_+^2(x_+)+O(z^3).
\end{equation}
The direct second-order truncation would grow as \(z^2\), whereas the exact logarithm grows at most linearly in \(z\) because \(0\le f(u)\le1\). 
We therefore damp the second-cumulant term by a factor \(1/(1+z)\), which preserves the expansion through \(O(z^2)\) and gives linear large-\(z\) growth.  
This gives
\begin{equation} \label{eq:quick_growth_factor}
	\mathcal G_+(H,x_+)\equiv \exp\left[ 2H\mu_+(x_+) + \frac{2H^2\sigma_+^2(x_+)}{1+2H} \right],
\end{equation}
with the threshold value understood as the continuous limit \(\mathcal G_+(0,0)=1\).

\subsubsection{Final closed-form formula}

From the low-density relation \(N_{\rm low}=N_*(\Gamma_0 L)^2\), 
the explicit prefactor Eq.~\eqref{eq:quick_Nstar_exact}, 
the unstable action Eq.~\eqref{eq:quick_Heff}, and the output factor Eq.~\eqref{eq:quick_growth_factor}, 
we obtain the quick estimate formula:
\begin{equation} \label{eq:quick_final_density}
	\langle N\rangle_{\rm vac}^{\rm quick,lin} = N_*(\Gamma_0 L)^2 \mathcal G_+\left(\frac{T_2}{L}\Phi(\Gamma_0 L),x_+\right).
\end{equation}
At large density the full simulation saturates, whereas the linear estimate does not. 
We therefore cap the estimate at the excited-molecule ceiling
\begin{equation} \label{eq:quick_Nsat}
	N_{\rm sat}=\rho_{ee}(0)nSL,
\end{equation}
We then take
\begin{equation} \label{eq:quick_final_capped}
	\langle N\rangle_{\rm vac}^{\rm quick} = \min\left( \langle N\rangle_{\rm vac}^{\rm quick,lin}, N_{\rm sat} \right).
\end{equation}
Here the auxiliary quantities are assembled from Eq.~\eqref{eq:Gamma_max},
\eqref{eq:hc_threshold}, \eqref{eq:quickelem_hdelta},
\eqref{eq:quick_Phi}, \eqref{eq:quick_tc}, \eqref{eq:quick_tau_out},
\eqref{eq:quickelem_x}, \eqref{eq:quick_Rout}, \eqref{eq:quick_sigma_explicit},
and \eqref{eq:quick_growth_factor}:
\begin{align}
	\Gamma_0 L                  & =\frac{\Omega}{2}\,|\alpha_{eg}\rho_{eg}(0)|\,nL,
	                   & \Gamma_0 L                                                      & =\frac{\pi}{2}\quad\text{at threshold},
	\label{eq:quick_g_of_n}                                                                                                                     \\[4pt]
	\delta             & =\Gamma_0 L-\frac{\pi}{2},
	                   & a                                                               & =1-\frac{\pi}{4},
	\label{eq:quickelem_aux1}                                                                                                                   \\[4pt]
	\Phi(\Gamma_0 L) & =\Theta(\delta)\left[
		\delta
		-
		\frac{1}{\Delta}
		\ln\!\left(
		\frac{(2a\delta+1-\Delta)(1+\Delta)}
		{(2a\delta+1+\Delta)(1-\Delta)}
		\right)
		\right],
	                   & H                                                               & =\frac{T_2}{L}\Phi(\Gamma_0 L),
	\label{eq:quick_Phi_final}                                                                                                                  \\[4pt]
	\Delta             & =\sqrt{\pi-3},
	                   & \tau_{\rm out}                                                  & =L\!\left(1+\frac{4}{\pi^2}\right),
	\label{eq:quick_H_final}                                                                                                                    \\[4pt]
	t_c                & =T_2\ln\!\left(\frac{\Gamma_0 L}{\pi/2}\right)\Theta(\delta),
	                   & x_+                                                             & =\frac{t_c}{\tau_{\rm out}},
	\label{eq:quick_tc_final}                                                                                                                   \\[4pt]
	\mu_+(x_+)         & =\Theta(x_+)\!\left[1-\frac{1-e^{-x_+}}{x_+}\right],
	                   &                                                                 & 
	\label{eq:quick_Rx_final}                                                                                                                   \\[4pt]
	\sigma_+^2(x_+)    & =\Theta(x_+)\!\left[
	1-\frac{2(1-e^{-x_+})}{x_+}
	+\frac{1-e^{-2x_+}}{2x_+}
	-\mu_+(x_+)^2\right],
	                   & \mathcal G_+(H,x_+)                                             & =\exp\!\left[
		2H\mu_+(x_+)+\frac{2H^2\sigma_+^2(x_+)}{1+2H}
		\right],
	\label{eq:quick_sigma_final}                                                                                                                \\[4pt]
	N_*                & =
	\frac{T_2}{8L^2}
	\left[
	2L
	-
	T_2 e^{-2(T_{\rm obs}-L)/T_2}
	+
	T_2 e^{-2T_{\rm obs}/T_2}
	\right],
	                   & N_{\rm sat}                                                     & =\rho_{ee}(0)\,n\,SL.
	\label{eq:quick_Nstar_final}
\end{align}

The compact notation used in the main text is obtained by identifying
\begin{equation} \label{eq:quick_maintext_S}
	\mathcal S\left(\Gamma_0L,\frac{T_2}{L}\right) \equiv \ln \mathcal G_+\left[ \frac{T_2}{L}\Phi(\Gamma_0L),x_+ \right].
\end{equation}
Here $n$ is the molecular number density, $L$ the medium length, $T_2$ the dephasing time, $T_{\rm obs}$ the observation window, 
$\Omega=\omega_{eg}/2$ the carrier frequency, $\alpha_{eg}$ the effective two-photon polarizability, $\rho_{eg}(0)$ the initial medium coherence, 
and $\Theta$ the Heaviside step function. 

For $\Gamma_0 L\le \pi/2$ one has $\Phi=0$ and the estimate reduces to the stable limit \(\langle N\rangle\propto (\Gamma_0 L)^2\propto n^2\).
For $\Gamma_0 L> \pi/2$ the output is exponentially amplified, with the exponent controlled by the approximate log-gain \eqref{eq:quick_Heff} and the output-formation closure \eqref{eq:quick_growth_factor}.

\end{document}